\documentclass[12pt, draftclsnofoot, onecolumn]{IEEEtran}
\usepackage[utf8]{inputenc}
\usepackage{color}
\usepackage{amsmath}
\usepackage{amssymb} 
\usepackage{graphicx}
\usepackage{xspace}
\usepackage{bbm} 
\usepackage{bbold}
\usepackage{comment}
\usepackage{color}
\usepackage{cite}

\newcommand{\Ex}[2]{\ensuremath{\mathbb{E}_{#1}\lefto[#2\right]}} 	% expectation
\newcommand{\lefto}{\mathopen{}\left}
\newcommand{\iid}{i.i.d.\@\xspace}

\providecommand{\indi}[1]{\mathbbm{1}\mathopen{}\left\{#1\right\}}

\usepackage{mathtools}

\newcommand{\E}[1]{\mathbb{E}\lefto[#1\right]}

\newcommand{\pr}[1]{\mathrm{Pr}\lefto[#1\right]}

\newcommand{\prBigg}[1]{\mathrm{Pr}\mathopen{}\Bigg[#1\Bigg]}

\newcommand{\unit}[1]{\ % works only outside mathmode or in flalign
    \ifmmode & \left[\textup{#1}\right] \hspace{1cm}
    \else \hfill $\left[\textup{#1}\right]$ \hspace{1cm}
    \fi
}
\newcommand{\vectt}[1]{%
  \if#1\relax\bm{#1}\else\mathbf{#1}\fi
}


\begin{document}

\title{5G Wireless Network Slicing for \\ eMBB, URLLC, and mMTC: \\ A Communication-Theoretic View}

\author{Petar Popovski$^{1}$~\IEEEmembership{Fellow,~IEEE}, Kasper F. Trillingsgaard$^{1}$~\IEEEmembership{Student~Member,~IEEE}, 
Osvaldo Simeone$^{2}$~\IEEEmembership{Fellow,~IEEE}, 
and Giuseppe Durisi$^{3}$~\IEEEmembership{Senior~Member,~IEEE} 
\IEEEcompsocitemizethanks{\IEEEcompsocthanksitem
$^{1}$ Department of Electronic Systems, Aalborg University, Denmark (email: \{petarp,kft\}@es.aau.dk).
\IEEEcompsocthanksitem
$^{2}$ King's College London, United Kingdom (email: osvaldo.simeone@kcl.ac.uk).
\IEEEcompsocthanksitem
$^{3}$ Department of Electrical Engineering, Chalmers University of Technology, Sweden (email: durisi@chalmers.se).
\IEEEcompsocthanksitem The work of Petar Popovski and Kasper F. Trillingsgaard has been in part supported by the European Research Council (Horizon 2020 ERC Consolidator Grant Nr. 648382 WILLOW).
% and the Horizon 2020 project ONE5G (ICT -
% 760809) receiving funds from the European Union. The authors would like to
% acknowledge the contributions of their colleagues in the ONE5G project, although the views expressed in this contribution are those of the authors and do not necessarily represent the project. 
The work of Osvaldo Simeone has received funding from the European  Research  Council (ERC) under the European Union Horizon 2020 research and innovation program (grant agreement 725731).
The work of Giuseppe Durisi was partly supported by the Swedish Research Council under grant 2016-03293.
}
}

\maketitle
\thispagestyle{plain}
\pagestyle{plain}

\begin{abstract}
The grand objective of 5G wireless technology is to support three generic services with vastly heterogeneous requirements: enhanced mobile broadband (eMBB), massive machine-type communications (mMTC), and ultra-reliable low-latency communications (URLLC). Service heterogeneity can be accommodated by 
network slicing, through which each service is allocated resources to provide performance guarantees and isolation from the other services. Slicing of the Radio Access Network (RAN) is typically done by means of orthogonal resource allocation among the services. This work studies the potential advantages of allowing for non-orthogonal sharing of RAN resources in uplink communications from a set of eMBB, mMTC and URLLC devices to a common base station. The approach is referred to as Heterogeneous Non-Orthogonal Multiple Access (H-NOMA), in contrast to the conventional NOMA techniques that involve users with homogeneous requirements and hence can be investigated through a standard multiple access channel. The study devises a communication-theoretic model that accounts for the heterogeneous requirements and characteristics of the three services. The concept of \emph{reliability diversity} is introduced as a design principle that leverages the different reliability requirements across the services in order to ensure performance guarantees with non-orthogonal RAN slicing. This study reveals that H-NOMA can lead, in some regimes, to significant gains in terms of performance trade-offs among the three generic services as compared to orthogonal slicing.
\end{abstract}

\section{Introduction}\label{sec:introduction}

During the past few years, there has been a growing consensus that 5G wireless systems will support three generic services, which, according ITU-R, are classified as enhanced mobile broadband (eMBB), massive machine-type communications (mMTC), and ultra-reliable and low-latency communications (URLLC) (also referred to as mission-critical communications) ~\cite{{ITUR,tr38802}}. A succinct characterization of these services can be put forward as follows: (\emph{a}) eMBB supports stable connections with very high peak data rates, as well as moderate rates for cell-edge users;
(\emph{b}) mMTC supports a massive number of Internet of Things (IoT) devices, which are only sporadically active and send small data payloads; (\emph{c}) URLLC supports low-latency transmissions of small payloads with very high reliability from a limited set of terminals, which are active according to patterns typically specified by outside events, such as alarms. This paper studies the problem of enabling the coexistence of the three heterogeneous services within the same Radio Access Network (RAN) architecture. We describe below in more details the requirements of the three services.   

eMBB traffic can be considered to be a direct extension of the 4G broadband service. It is characterized by large payloads and by a device activation pattern that remains stable over an extended time interval. 
This allows the network to schedule wireless resources to the eMBB devices such that no two eMBB devices access the same resource simultaneously. 
The objective of the eMBB service is to maximize the data rate, while guaranteeing a moderate reliability, with packet error rate (PER) on the order of $10^{-3}$.

In contrast, an mMTC device is active intermittently and uses a fixed, typically low, transmission rate in the uplink. A huge number of mMTC devices may be connected to a given base station (BS), but at a given time only an unknown (random) subset of them becomes active and attempt to send their data. 
The large number of potentially active mMTC devices makes it infeasible to allocate 
\emph{a priori} resources to individual mMTC devices. Instead, it is necessary to provide resources that can be shared through random access. The size of the active subset of mMTC devices is a random variable, whose average value measures the mMTC traffic arrival rate. 
The objective in the design of mMTC is to maximize the arrival rate that can be supported in a given radio resource.
The targeted PER of an individual mMTC transmission is typically low, e.g., on the order of $10^{-1}$.

Finally, URLLC transmissions are also intermittent, but the set of potential URLLC transmitters is much smaller than for mMTC. 
Supporting intermittent URLLC transmissions requires a combination of scheduling, so as to ensure a certain amount of predictability in the available resources and thus support high reliability; as well as random access, in order to avoid that too many resources being idle due to the intermittent traffic. 
Due to the low latency requirements, a URLLC transmission should be localized in time. Diversity, which is critical to achieve high reliability~\cite{ferrante18-03a}, can hence be achieved only using multiple frequency or spatial resources. 
The rate of a URLLC transmission is relatively low, and the main requirement is ensuring a high reliability level, with a PER typically lower than $10^{-5}$, despite the small blocklengths.

%While there has been a significant research on how to design each individual service \cite{5GTutorialOverview,TowardsmMTC,PPURLLC}, mechanisms to support more than one service within a 5G wireless systems have only recently started to be addressed, see \cite{NetworkSlicing5G,anand2017joint}.  The central premise of this work is that the requirements and the characteristics of the three generic services are \emph{heterogeneous} and w
In 5G, heterogeneous services are allowed to coexist within the same network architecture by means of \emph{network slicing} \cite{NetworkSlicing5G}. Network slicing allocates the network computing, storage, and communication resources among the active services with the aim of guaranteeing their isolation and given performance levels. In this paper, we are interested in the ``slicing" of RAN communication resources for wireless access. The conventional approach to slice the RAN is to allocate orthogonal radio resources to eMBB, mMTC, and URLLC devices in time and/or frequency domains, consistently with the orthogonal allocation of wired communication resources. However, wireless resources are essentially different due to their shared nature. Using communication-theoretic analysis, this work demonstrates  that a non-orthogonal allocation that is informed by the heterogeneous requirements of the three services can outperform the standard orthogonal approach. Importantly, the considered non-orthogonal approach multiplexes heterogeneous services, and is hence markedly distinct from the conventional Non-Orthogonal Multiple Access (NOMA) methods that share radio resources only among devices of the same type (see, e.g., \cite{NOMA2018}). This is further discussed next.

\subsection{Network Slicing of Wireless Resources: H-OMA and H-NOMA}

\begin{figure}[t!]
\centering
  \includegraphics[width=8.3cm]{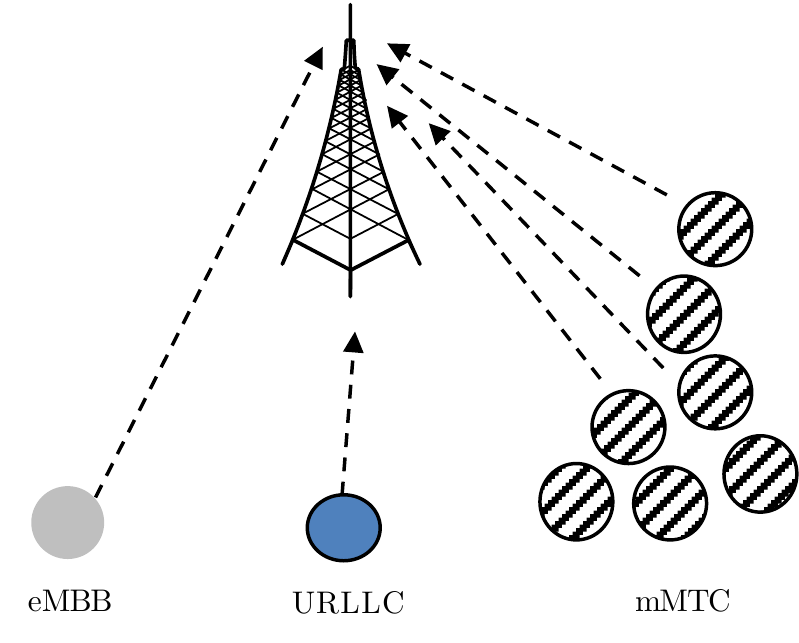}
  \caption{The considered scenario with uplink transmissions to a common base station (BS) from devices using the three generic 5G services.}
 \label{fig:BSuplink}
\end{figure}

\begin{figure}[t!]
\centering
  \includegraphics[width=8.3cm]{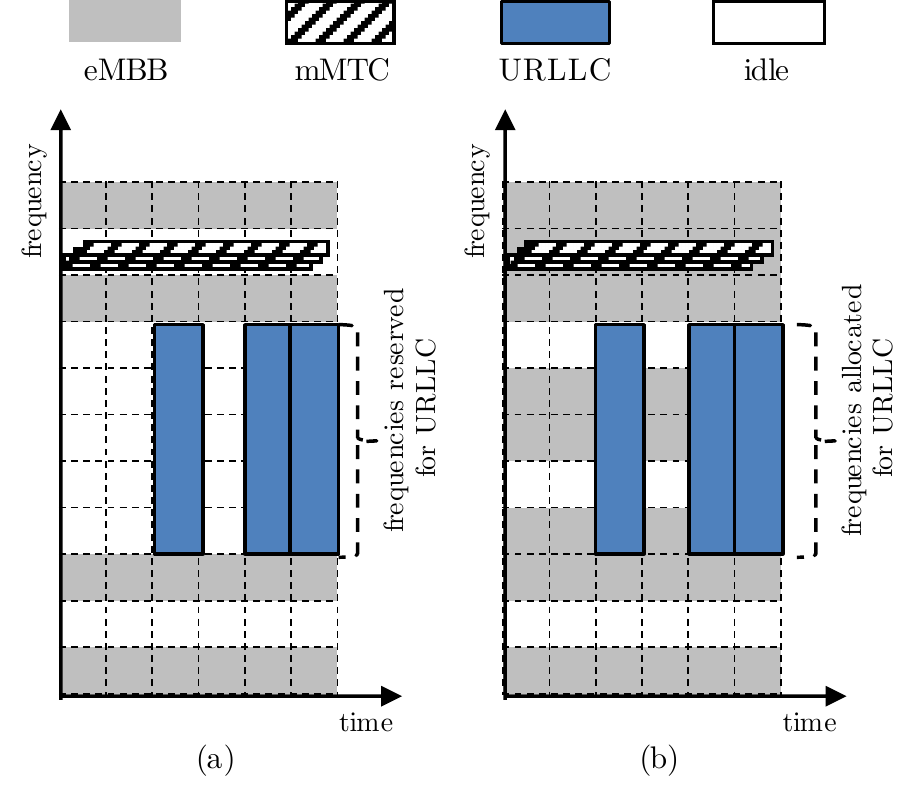}
  \caption{Illustration of the slicing of the wireless resources in a time-frequency frame for supporting the three generic services with: (a) Heterogeneous Orthogonal Multiple Access (H-OMA) (b) Heterogeneous Non-Orthogonal Multiple Access (H-NOMA). The idle time-frequency blocks are not used for transmission due to absence of traffic. With H-OMA, some of the frequency channels are reserved to URLLC traffic, whereas with H-NOMA the same channels are allocated to both URLLC and eMBB.}    
 \label{fig:BasicSlicingPic}
\end{figure}

Consider an uplink scenario in which a set of eMBB, mMTC and URLLC devices is connected to a common BS, as shown in Fig.~\ref{fig:BSuplink}. 
We note that the designing uplink access is more complex than the corresponding problem for the downlink due to the lack of coordination among users. Orthogonal and non-orthogonal slicing of the RAN among the three services are illustrated in Fig. \ref{fig:BasicSlicingPic}(a) and (b), respectively. 

The conventional orthogonal allocation depicted in Fig.~\ref{fig:BasicSlicingPic}(a) operates in the frequency domain and allots different frequency channels to eMBB, mMTC, or URLLC devices. eMBB and mMTC transmissions are allowed to span multiple time resources. In contrast, in order to guarantee the latency requirements discussed above, URLLC transmissions are localized in time and are spread over multiple frequency channels to gain diversity. Furthermore, since the URLLC traffic is bursty, the resources allocated to URLLC users may be largely unused. This is because the channels reserved for URLLC are idle in the absence of URLLC transmission. 

Importantly, orthogonal slicing does not preclude the sharing of wireless resources among devices of the same type. For example, multiple eMBB users may transmit on the same allotted frequency channels by using NOMA \cite{NOMA2018}. Therefore, in order to distinguish orthogonality among signals originating from devices of the same type, as in conventional Orthogonal Multiple Access (OMA), from the orthogonality among different services, we refer to the approach in Fig.~\ref{fig:BasicSlicingPic}(a) as  \emph{Heterogeneous Orthogonal Multiple Access (H-OMA)}. 

As mentioned, in this work, we investigate the potential advantages of a non-orthogonal allocation of RAN resources among multiple services, which we refer to as \emph{Heterogeneous Non-Orthogonal Multiple Access (H-NOMA).} Fig.~\ref{fig:BasicSlicingPic}(b) depicts an instance of H-NOMA. By comparison with the H-OMA solution in Fig. \ref{fig:BasicSlicingPic}(a), under H-NOMA, the frequency resources that were allocated only to mMTC or URLLC traffic can also be granted to the eMBB users. In this way, H-NOMA may allow for a more efficient use of radio resources as compared to H-OMA by avoiding unused resources due to URLLC or mMTC inactivity. This may yield a higher spectral efficiency for the eMBB users that can benefit from the intermittent nature of mMTC and URLLC traffic. However, the mutual interference between eMBB and mMTC or URLLC transmissions may significantly degrade the performance for all the involved services. Ensuring desired performance levels is hence more challenging with H-NOMA. 

%However, while the isolation of the slices for HOMA is evident, it is less so for HNOMA. Another key contribution of our work is analysis of the ways in which service isolation can be achieved with non-orthogonal slicing. 

\begin{figure}[t!]
\centering
  \includegraphics[width=8.3cm]{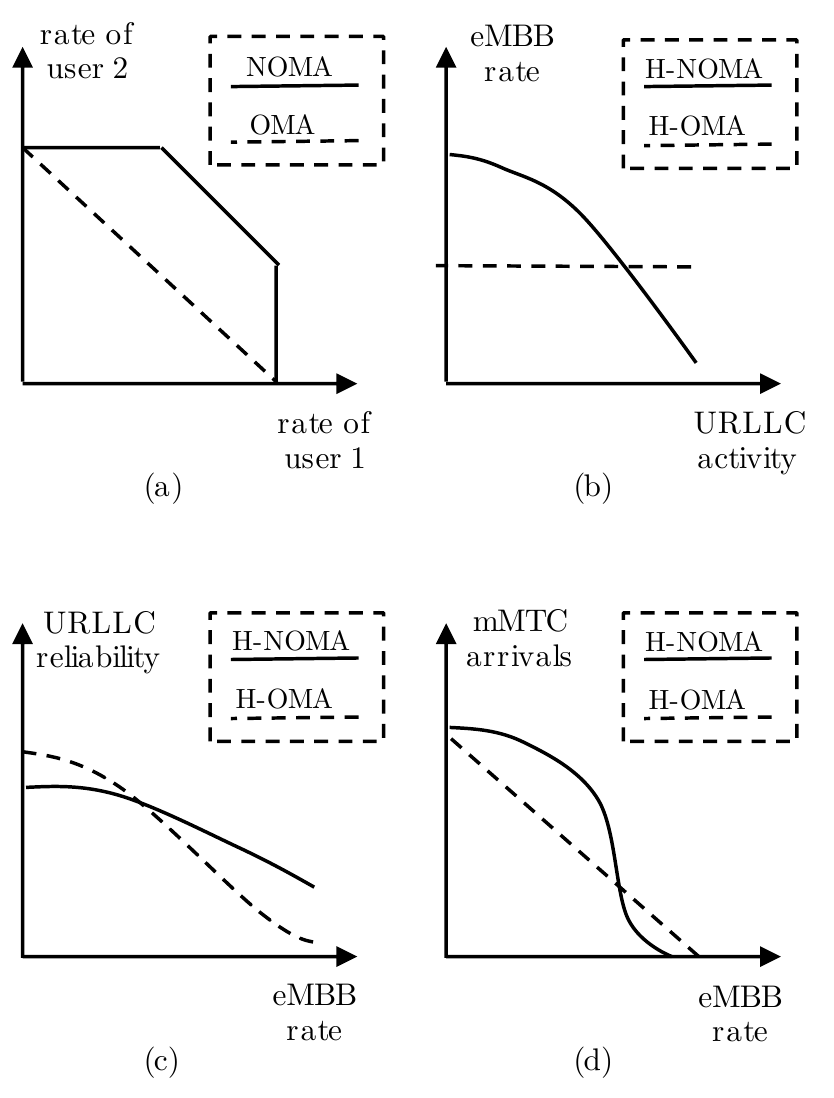}
  \caption{Illustration of performance trade-offs for: (a) standard OMA and NOMA within the same traffic type; (b,c,d) H-OMA and H-NOMA between heterogeneous services.}
 \label{fig:ResultTypes}
\end{figure}
In this paper, we tackle this problem by developing a communication-theoretic model that aims at capturing the essential performance trade-offs and design insights for H-OMA and H-NOMA. More specifically, the main goals of this work can be illustrated using Fig.~\ref{fig:ResultTypes}, as discussed next. 

To start, Fig.~\ref{fig:ResultTypes}(a) depicts the type of results that are of interest when studying conventional OMA and NOMA within a given service type, as done in a growing line of work \cite{dai2015non,NOMA2018}. These results rely on the classical analysis of the multiple access channel, in which all users have identical reliability requirements and block lengths and the goal is to characterize the region of achievable rates \cite{cover2012elements} as the block length grows large. 

%Recall that the classical MAC model is based on the assumption that all channel uses are statistically identical, while the characteristics and the requirements of the users are homogeneous. The classical MAC model can capture the tradeoff among the asymptotically achievable rates of different competing users. However, it cannot capture the features of the heterogeneous services, such as different activation patterns or reliability requirements. 

In contrast, Fig.~\ref{fig:ResultTypes}-(b,c,d) exemplify the type of results that are of interest when evaluating the performance trade-offs between heterogeneous services that are allowed by H-OMA and H-NOMA. As a first example to be further elaborated on in the paper, Fig.~\ref{fig:ResultTypes}(b) shows the trade-off between the URLLC activity, i.e., the probability of URLLC devices being active, and the eMBB transmission rate or spectral efficiency. The figure illustrates the fact that the eMBB rate is not affected by the URLLC packet arrival rate under H-OMA, while the resulting interference impairs the performance of H-NOMA. As an alternative performance evaluation, Fig.~\ref{fig:ResultTypes}(c) shows the trade-off between the reliability of URLLC transmissions and the eMBB rate. The example highlights the fact that a non-trivial trade-off exists for both H-OMA and H-NOMA. In fact, URLLC reliability can be improved by taking away frequency resources from eMBB and allocating them to URLLC. As a final illustration, Fig.~\ref{fig:ResultTypes}(d) depicts the trade-off between the arrival rate of mMTC devices and the eMBB rate. In a manner similar to Fig. ~\ref{fig:ResultTypes}(c), this figure suggests that, even under H-OMA, the spectral efficiency of the eMBB user that shares the resources with mMTC devices can be traded off for the mMTC arrival rate by a proper allocation of radio resources.

\subsection{Further Related Works}

In addition to the mentioned literature on conventional NOMA, here we briefly review works that directly tackle the coexistence of heterogeneous services. A logical architecture for network slicing in 5G in the presence of orthogonal slicing has been presented in \cite{NetworkSlicing5G} and ~\cite{ServiceOrientedSlicing}. The downlink multiplexing of URLLC and eMBB is studied in \cite{SignalSpaceDiversityeMBBURLLC} and \cite{anand2017joint}. These works investigates the dynamic scheduling of URLLC traffic over ongoing eMBB transmissions by abstracting the operation at the physical layer. In \cite{Globecom17URLLCmMTC}, the authors treat the problem of resource allocation for mMTC and URLLC in a new radio (NR) setting by focusing on the role of feedback. Orthogonal resource allocation for mMTC and eMBB users is studied in \cite{MahmoodeMBBmMTC} by accounting for inter-cell interference. In \cite{NOMA5GChinaComm}, grant-free uplink transmissions are considered for the three services by considering concrete transmission/modulation/spreading methods for supporting the three services.

\subsection{Main Contributions}
The main contributions are as follows.
\begin{itemize}
\item We propose a 
communication-theoretic model that is tractable and yet captures the key features and requirements of the three services. Unlike \cite{NOMA5GChinaComm}, in which the authors focus on grant-free access for all services, the proposed model takes into account the difference in arrival processes and traffic dynamics that are inherent to each individual service. 
%The model also accounts for different decoding architectures at the BS.
The proposed model can be seen as an extension of the classical multiple access channel model that underlies the analysis of conventional NOMA in the sense that it accounts for the coexistence of heterogeneous services.
\item We first analyze the performance of orthogonal slicing, or H-OMA, for all three services. We focus on achievable transmission rates for eMBB and URLLC, under the respective target reliability, and on the throughput for mMTC.
\item We then consider the performance of H-NOMA. Although the modeling approach allows to study an arbitrary combination of services, in this paper we have focused on the analysis of two specific cases as illustrated in Fig.~\ref{fig:BasicSlicingPic}, namely: (\emph{i}) slicing for URLLC and eMBB, and (\emph{ii}) slicing for mMTC and eMBB. In the case of URLLC-eMBB slicing, among other schemes, we consider the technique of puncturing, which is currently under consideration in 3GPP~\cite{anand2017joint}. It is noted that, while of interest, H-NOMA between URLCC and mMTC may be problematic due to the need to ensure reliability guarantees for URLLC devices in the presence of the random interference patterns caused by mMTC transmissions. 
\item Among the main conclusions, our study demonstates that non-orthogonal slicing, or H-NOMA, can achieve service isolation in the sense of ensuring performance levels for all services by leveraging their heterogeneous reliability requirements. 
We refer to this design principle as \emph{reliability diversity}. As it will be discussed, the heterogeneity leveraged by reliability diversity is not only in terms of the numerical values of the reliability levels, but also in terms of very definition of reliability across the three services.
%This refers to discussed widely different reliability levels, but also to the distinct definitions of reliability for the three services. 
For example, the reliability metric typically considered for mMTC is the fraction of detected devices among the massive set of active users, whereas for eMBB and URLLC services one typically adopts the classical frame error rate. 
Our results show that, if reliability diversity is properly exploited, non-orthogonal slicing can lead, in some regimes, to important gains in terms of performance trade-offs among the three generic services. 
\end{itemize}

The paper is organized as follows. The next section presents the system model and provides a performance analysis of each of the three services when considered in isolation. Section~\ref{sec:eMBBURLLC} treats the slicing of resources to support eMBB and URLLC, while Section~\ref{sec:eMBBmMTC} is dedicated to the slicing of resources for eMBB and mMTC.  Both sections provide  a description of the proposed theoretical framework as well as numerical results illustrating the tradeoff between the services for both H-OMA and H-NOMA schemes.
The conclusions are given in Section~\ref{sec:conclusion}, while Section~\ref{sec:extensions} contains  discussion on possible generalizations of the model considered in this paper.
Two appendices, containing the technical details of some of the derivations, conclude the paper.

\section{System Model}

\begin{figure}[t!]
\centering
  \includegraphics[width=8.3cm]{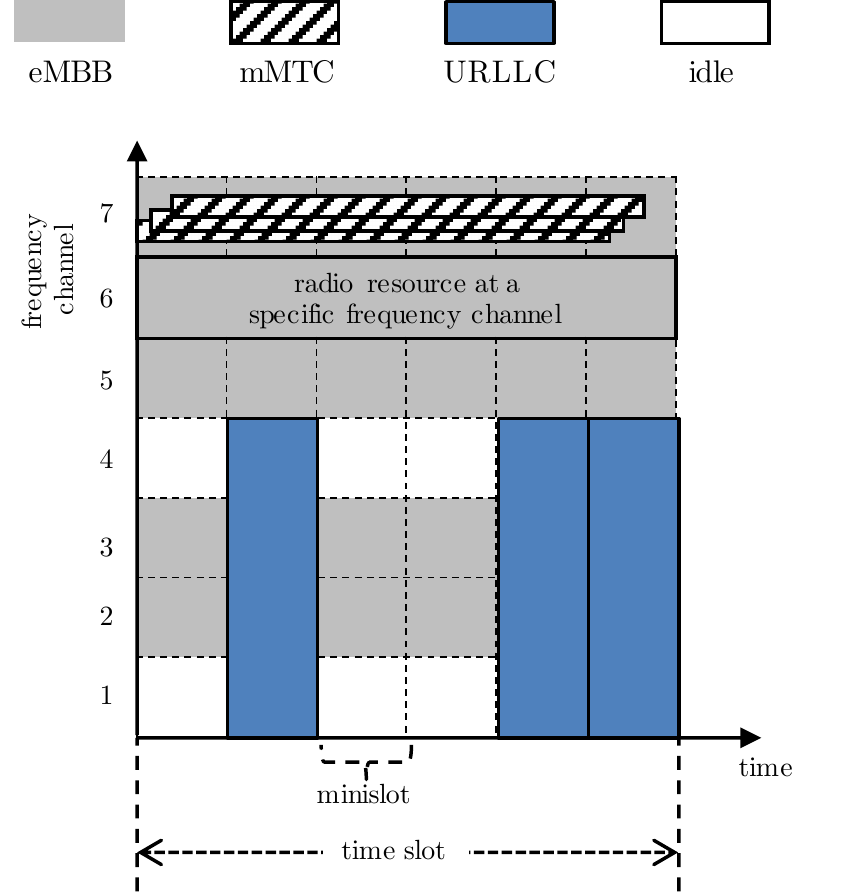}
  \caption{An example of H-NOMA allocation in the time-frequency grid with $F=7$ resources and $S=6$ minislots. A single resource (frequency channel) is allocated for mMTC transmission. Each URLLC transmission is spread over $F_U=4$ frequency channels.}
 \label{fig:TFmodel}
\end{figure}

We are interested in understanding how the three service described in Section~\ref{sec:introduction}, i.e., eMBB, URLLC, and mMTC, should efficiently share the same radio resources in the uplink when communicating to a common BS. 
We consider $F$ radio resources, where each resource occupies a single frequency channel and a single time slot. A radio resource, which is indexed by $f\in \{1,...,F\}$, contains $n$ symbols. 
The $n$ symbols are further divided into $S$ minislots, where each minislot consists of $n_S =n/S$ symbols. Fig.~\ref{fig:TFmodel} shows an example of a time-frequency grid.  

We assume that the transmission of an eMBB user occupies a single radio resource at a given frequency $f\in \{1,...,F\}$. 
In contrast, due to latency constraints, a URLLC user transmits within a single minislot across a subset of $F_U\leq F$ frequency channels. An URLLC device may be active in an allocated minislot with probability $a_U$.
Finally, the set of mMTC users is allowed to access the channel only at a specified radio resource $f_M$; for the example on Fig.~\ref{fig:TFmodel} we have $f_M=F$. 
The number $A_M$ of active mMTC devices in such a resource is distributed  as $A_M \sim \text{Poisson}(\lambda_M)$, where  $\lambda_M$ is the mean value, referred to as mMTC arrival rate. 

Some comments  regarding the modelling choices made above are in order. 
First, for eMBB traffic, we focus on the standard scheduled transmission phase, hence assuming that radio access and competition among eMBB devices have been resolved prior to the considered time slot.
Second, we do not model collisions among URLLC devices.
We assume instead that a single URLLC device is allocated a number of minislots in the given slot, over which it is active with some probability. 
On the contrary, we do model the random access phase for mMTC traffic, since this is the key transmission phase for this type of traffic, due to the massive population of devices. 
Extensions of our model will be discussed in Section~\ref{sec:extensions}. 

Each radio resource $f$ is assumed to be within the time- and frequency-coherence interval of the wireless channel, so that the wireless channel coefficients are constant within each radio resource. 
Furthermore, we assume that the channel coefficients fade independently across the $F$ radio resources. The channel coefficients of the eMBB, URLLC, and the mMTC devices, which we denote by $H_{B,f}$, $H_{U,f}$, and $H_{m,f}$, $m \in \{0,\ldots,A_M\}$,\footnote{Throughout, we use the convention that the subscripts $B$, $U$, and $M$ indicate a quantity referring to eMBB, URLLC, and mMTC, respectively.}  are independent and Rayleigh distributed, i.e., $H_{B,f} \sim \mathcal{CN}(0,\Gamma_M) $, $H_{U,f} \sim \mathcal{CN}(0,\Gamma_U)$, and $H_{m,f} \sim \mathcal{CN}(0,\Gamma_M)$ for $m \in \{0,\ldots,A_M\}$ across all radio resources $f\in{1,...,F}$. 
The channel gains for the three services in a radio resource $f$ are denoted by $G_{B,f} = |H_{B,f}|^2$, $G_{U,f} = |H_{U,f}|^2$, and $G_{m,f} = |H_{m,f}|^2$ for $m\in{1,...,A_M}$.

The average transmission power of all devices is normalized to one. 
The differences in the actual transmission power across various users and in the path loss are  accounted for through the average channel gains $\Gamma_B$, $\Gamma_U$, and $\Gamma_{M}$. 
Furthermore, the power of the noise at the BS is also normalized to one, so that the received power equals the signal-to-noise ratio (SNR) for each device. The number of symbols $n_S$ in a minislot is assumed sufficiently large to justify an asymptotic information-theoretic analysis.
Extensions of our analysis to capture finite-blocklength effects~\cite{GDTKPP} will be considered in future works. Due to latency and protocol constraints to be detailed later, no channel-state information (CSI) is assumed at the URLLC and at the mMTC devices. 
In contrast, the eMBB devices are assumed to have perfect CSI. 
Finally, the BS is assumed to have perfect CSI.

% to On the other hand, the eMBB devices are allowed to adapt the transmission power based on the realized channel coefficient using a power adaptation policy $\mathbb{p}_B(\cdot)$ that needs to satisfy the constraint $\E{\mathbb{p}_B(G_{B,f})} \leq 1$.

The error probabilities of the eMBB, URLLC, and mMTC devices are denoted as $\mathrm{Pr}(E_B), \mathrm{Pr}(E_U)$ and $\mathrm{Pr}(E_M)$, respectively. 
These probabilities must satisfy the reliability requirements $\mathrm{Pr}(E_B) \leq \epsilon_B$, $\mathrm{Pr}(E_U) \leq \epsilon_U$ and $\mathrm{Pr}(E_M) \leq \epsilon_M$, 
where
% According to the general expected properties of the three services, we assume that the reliability requirements satisfy the inequalities
\begin{equation} \label{eq:relcon} \epsilon_U \ll \epsilon_B\ll \epsilon_M.\end{equation}
% Details on the assumed operation for each service will be introduced later in this section. 
The large differences in reliability levels among the services, as well as their different definitions, which we will introduce shortly, motivate the introduction of the concept of \emph{reliability diversity}. 
Reliability diversity  refers to system design choices that leverage the differences among the supported services in terms of reliability requirements and definitions. For example, as we will see, strict per-packet reliability guarantees are typically enforced for eMBB and URLLC devices, whereas the notion of reliability for mMTC devices is less stringent and typically involves the computation of averages over a large group of active devices.
%As we will see, reliability diversity can be used to guide the design of non-orthogonal slicing solutions. 

%However, as discussed in Section~\ref{sec:DiscussionGeneralSlicing}, the methodology developed here can be used to study those cases as well.

\subsection{Signal Model}

To summarize the main assumptions discussed so far and to fix the notation, we assume that each eMBB user is scheduled on a single frequency channel within the considered $F$ frequency resources; each URLLC device occupies $F_U \leq F$ frequencies resources, numbered without loss of generality as $f=1, \ldots, F_U$, in a given minislot; and a set of mMTC devices is available for transmission in a channel frequency $f_M\in\{1,\ldots,F\}$.

Let $\mathbf{Y}_{s,f}\in\mathbb{C}^{n_S}$ 
denote the received vector corresponding to the minislot $s \in\{1, 2, \ldots, S\}$ and the frequency channel $f \in \{1, 2, \ldots F\}$. Based on the given assumptions, the received signal can be written as
\begin{IEEEeqnarray}{rCL}\label{eq:rxsignal}
\mathbf{Y}_{s,f}&=&H_{B,f}\mathbf{X}_{B,s,f}  + H_{U,f} \mathbf{X}_{U,s,f} \nonumber \\ 
&&+\sum_{m=1}^{{A_M}}  H_{[m],f} \mathbf{X}_{[m],s,f} +\mathbf{Z}_{s,f}, 
\end{IEEEeqnarray} 
where $\mathbf{X}_{B,s,f}$  is the signal transmitted by an eMBB user scheduled in the frequency resource $f$; $\mathbf{X}_{U,s,f}$ is the signal transmitted by a URLLC device transmitting in minislot $s$ and frequency $f$; $\mathbf{X}_{[m],s,f}$ is the signal transmitted by one of the $A_M$ active mMTC devices in frequency $f$; and $\mathbf{Z}_{s,f}$ represents the noise vector, whose entries are i.i.d. Gaussian with zero mean and unit variance. The notation $[m]$ for the mMTC devices, which indicates ordering, will be formally introduced in Section \ref{subsec:mmtc}. 

We emphasize that the transmitted eMBB signal $\mathbf{X}_{B,s,f}$ in~\eqref{eq:rxsignal} is zero if no eMBB user is scheduled in frequency channel $f$; similarly, the URLLC signal $\mathbf{X}_{U,s,f}$ is zero if no URLLC device transmits in minislot $s$ and frequency $f$, e.g., if $f>F_U$; and the mMTC signals $\{\mathbf{X}_{[m],s,f}\}$ are similarly all equal to zero if the channel $f$ is not allocated to mMTC traffic, i.e., if $f\neq f_M$.

As discussed, with H-OMA, resources are allocated exclusively to one of the three services, while, with H-NOMA, resources can be shared. In the remainder of this section, we study the performance of the three traffic types in an H-OMA setting, that is, in the absence of mutual interference. We also introduce the metrics that will be used to evaluate the performance of the three services.
% \begin{align}
% \mathbf{Y}_{s,f}=H_{B,f}\mathbf{X}_{B,s,f} \delta_{B,f} + H_{U,f} \mathbf{X}_{U,s,f} \zeta_{U,s} \delta_{U,f}+ \nonumber \\ 
% \delta_{M,f}\sum_{m=1}^{{A_M}}  H_{[m],F} \mathbf{X}_{[m],s} +\mathbf{Z}_{s,f} 
% \end{align}
% Here $\delta_{B,f}=1$ if an eMBB transmission is scheduled on the $f-$th channel and is $0$ otherwise. 
% w
% The indicator $\delta_{U,f}=1$ if $f\leq F_U$ and is $0$ otherwise.  The indicator $\zeta_{U,s}=1$ if there is a URLLC transmission in the $s-$th minislot and is $0$ otherwise; 
% The indicator $\delta_{M,f}=1$ if $f=f_M$ and is $0$ otherwise. 

\subsection{eMBB} \label{sec:eMBB}
% eMBB are allocated $F_B$ radio resources out of $F$ channels.  

% Without loss of generality, we assume that the channels allocated to eMBB are $\{f\}_{F-F_B+1}^F$.
% Then there are $F_B$ different eMBB devices and each eMBB device is preallocated to the channel where it is allowed to transmit. If there are more than $F_B$ eMBB devices in the system, they use the $F_B$ channels in a time-division manner. On the other hand, the setup is not the most general one, as it does not cater for the situation in which a single eMBB device allocates its power/rate to multiple frequency channels. Still, the used model allows us to obtain insights into the interaction among the three services and the impact of a more generalized model will be discussed in Section~\ref{sec:Discussion}.

% For Each eMBB device is given a single unique radio resource $f$. 
% As it will be discussed in the next sections, this model will allow to capture the key elements of the interaction among the three services. Possible generalization of the setup will be covered in Section~X.
Consider a radio resource $f$ allocated exclusively to an eMBB user. As mentioned, the eMBB is aware of the CSI $G_{B,f}$ and can use it in order to select its transmission power $P_B(G_{B,f})$. 
The objective is to transmit at the largest rate $r_{B,f}$ that is compatible with the outage probability requirement $\epsilon_B$ under a long-term average power constraint. This can be formulated as the optimization problem
\begin{IEEEeqnarray}{rCL}
  %\IEEEeqnarraymulticol{3}{l}{...}
  % a & = & b +c
  \mathrm{maximize} &\quad& r_{B}  \nonumber\\
  \mathrm{subject\, to} &\quad& \pr{\log_2\lefto(1+G_{B,f}P_B(G_{B,f})\right)< r_{B}}\leq \epsilon_B \notag \\
  &\quad& \Ex{}{P_B(G_{B,f})}=1.\label{eq:optprotp}
\end{IEEEeqnarray}
The optimal solution to this problem is given by truncated power inversion~\cite{caire1999optimum}. Accordingly, the eMBB device chooses a transmission power that is inversely proportional to the channel gain $G_{B,f}$ if the latter is above a given threshold $G^{\mathrm{min}}_{B,f}$, while it refrains from transmitting otherwise. 

Beside being theoretically justified by the mentioned rate-maximization problem, the threshold-based transmission strategy discussed above also captures the fact that eMBB devices only transmit if the current SNR is sufficient to satisfy minimal rate requirements. This is the case in most communication standards, such as LTE, in which the transmission mode is selected from a set of allowed modulation and coding schemes with given SNR constraints. 
As we will discuss below, the scheme has the additional analytical advantage of relating directly outage probability and probability of activation for an eMBB user. 
We remark that the analysis could be extended to other design criteria such as the maximization of the average transmission rate. 

% We note that If eMBB device adapts its rate to the current channel conditions, then it should use an adaptive rate. Nevertheless, there will still be a threshold $\eta_f$ below which no transmission is made. 
% In that sense, this alternative setup is analogous to the model used here. However, if the rate was adapted to the channel conditions, then $\beta_f$ would vary, which would complicate the analysis. 
% Such an investigation is left for future extensions of the model.

Based on the discussion above, the probability that the eMBB user transmits is given by
\begin{IEEEeqnarray}{rCl}\label{eq:actpB}
  a_B= \pr{G_{B,f} \geq G^{\mathrm{min}}_{B,f}} = e^{-G^{\mathrm{min}}_{B,f}/\Gamma_{B}}.
\end{IEEEeqnarray}
Furthermore, in the absence of interference from other services, the only source of outage for an eMBB transmission is precisely the event that an eMBB does not transmit because of an insufficient SNR level. 
Hence, the probability of error equals 
\begin{equation}\label{eq:eMBBreliab}
\mathrm{Pr}(E_B)=1-a_B.
\end{equation}
Imposing the reliability condition
\begin{equation}\label{eq:eMBBorthconstraint}
\mathrm{Pr}(E_B)=\epsilon_B
\end{equation}
we obtain the value of the threshold SNR
\begin{equation}\label{eq:Gminexplicit}
G^{\mathrm{min}}_{B,f}=\Gamma_B \ln\lefto(\frac{1}{1-\epsilon_B}\right).
\end{equation}
Note that, in the absence of interference and under the given assumptions, the threshold SNR $G^{\mathrm{min}}_{B,f}$ does not depend on the frequency channel $f$. This dependence is kept here in view to the extension to H-NOMA in the next sections.

Based on the power-inversion scheme, the instantaneous power $P_B(G_{B,f})$ is chosen as a function of the instantaneous channel gain $G_{B,f}$ as
\begin{IEEEeqnarray}{rCl}
  P_B(G_{B,f}) = \left\{\begin{array}{ll}
  \frac{G^{\mathrm{tar}}_{B,f}}{G_{B,f}} & \text{ if } G_{B,f} \geq G^{\mathrm{min}}_{B,f} \\
  0 & \text{ if } G_{B,f}< G^{\mathrm{min}}_{B,f},
  \end{array}
    \right.\label{eq:channel_inversion}
\end{IEEEeqnarray}
where $G^{\mathrm{tar}}_{B,f}$ is the target SNR, which is obtained from the threshold $G^{\mathrm{min}}_{B,f}$ by imposing the average power constraint as
\begin{IEEEeqnarray}{rCl}
1=\Ex{}{P_B(G_{B,f})}=\int_{G^{\mathrm{min}}_{B,f}}^\infty \frac{1}{\Gamma_{B}} e^{-x/\Gamma_{B}} P_B(x) \mathrm{d}x = \frac{G^{\mathrm{tar}}_{B,f}}{\Gamma_{B}} \gamma\mathopen{}\Big(0,\frac{G^{\mathrm{min}}_{B,f}}{\Gamma_{B}}\Big),
\end{IEEEeqnarray}
with $\gamma(\cdot,\cdot)$ being the lower incomplete gamma function. This implies that the target SNR is 
\begin{IEEEeqnarray}{rCl}
  G^{\mathrm{tar}}_{B,f} = \frac{\Gamma_{B}}{ \gamma\mathopen{}\Big(0,\frac{G^{\mathrm{min}}_{B,f}}{\Gamma_{B}}\Big)}. \label{eq:BetaFeta}
\end{IEEEeqnarray}

It follows from~\eqref{eq:actpB}--\eqref{eq:BetaFeta} that the solution to the problem (\ref{eq:optprotp}), which is the outage rate $r_{B,f}$ under outage probability $\epsilon_B$, is given by
% Sometimes, when the dependency on $\eta_f$ is clear from the context, we shall denote $\beta(\eta_f)$ and $q_B(\eta_f)$ by $\beta_f$ and $q_{B,f}$, respectively.
%
%
% We will characterize the eMBB transmission strategy through the threshold $G^{\mathrm{min}}_{B}$, as this uniquely determines the av. In particular, $\beta(\cdot)$ and $q_B(\cdot)$ are increasing and decreasing functions, respectively. 
%
% The performance metric of choice for eMBB traffic is the outage rate under probability of error $\epsilon_B$, which is given by
\begin{equation} \label{eq:Maximal_rB}
r_{B,f}=\log_2\bigl(1+G^{\mathrm{tar}}_{B,f}\bigr), \qquad \mathrm{[bits/symbol]}.
\end{equation}
%where the target SNR $G^{\mathrm{tar}}_{B,f}$ is obtained from (\ref{eq:BetaFeta}). 
% Overall, given the outage probability $\epsilon_B$, the eMBB rate $r_{B,f}$ is obtained from (\ref{eq:Maximal_rB}) by solving equations (\ref{eq:eMBBorthconstraint}) and (\ref{eq:BetaFeta}) in turn.
We refer to the resulting rate as $r^{\text{orth}}_B$ for reference. 
Note that it does not depend on $f$ under the assumptions of this section.
%If there are multiple eMBBs, we will be interested in the sum-rate across all eMBBs, denoted by $r^{\mathrm{orth}}_{B,\mathrm{sum}}$.

%he metric for evaluating eMBB should reflect how effective does the eMBB transmissions use the $F$ available channels. Assume that eMBB transmissions are allowed only on a subset ${\cal F}_B$ of all $F$ channels, then the metric of interest is:
% \begin{align}
% \rho_F=\frac{\sum_{f \in {\cal F}_B}r_{B,f}}{F}
% \end{align}
% In the special case where eMBB is using all $F$ channels and there are no resources allocated for URLLC/mMTC, the rate in each channel is chosen according to (\ref{eq:Maximal_rB}) and is identical for all $f$. In that case, $\rho_F=\log_2(1+\beta^{(\epsilon_B)})$.   

\subsection{URLLC} \label{sec:URLLC}

The URLLC device transmits data in the allocated $F_U$ frequency channels of a minislot, with activation probability $a_U$. 
Hence, the number of URLLC transmissions during the time slot is a random variable $S_U \sim \text{Bin}(S, a_U)$. 
We assume that each URLLC transmission carries a different message, and that, due to the low latency requirement, each message must be decoded as soon as the relevant minislot is received.
This implies that the URLLC device cannot code across multiple minislots.

Unlike eMBB users, the URLLC device is not aware of the CSI $\{G_{U,f}\}$ for the $F_U$ allocated frequency resources. This assumption is justified by the fact that CSI at the URLLC device would require signaling exchange before transmission, which entails extra latency as well as a potential loss in terms of reliability. In fact, the high reliability constraint would enforce an even higher reliability requirement on the auxiliary procedure of CSI signaling. 
As a result of the lack of CSI, no power or rate adaptation is possible for URLLC devices. 

We choose the rate $r_U$ as the performance metric of choice. 
In the absence of interference from other services, 
outage occurs with probability 
\begin{equation} \label{URLLCperformance}
\mathrm{Pr}(E_U)=\Pr\lefto(  \frac{1}{F_U} \sum_{f=1}^{F_U} \log_2 \left(1+G_{U,f}\right) <r_U
\right).
\end{equation}
Imposing the reliability condition $\mathrm{P}(E_U)=\epsilon_U$ allows us to obtain the maximum allowed rate~$r_U$. 
We will refer to this quantity as $r^{\text{orth}}_U(F_U)$ for reference. Note that increasing $F_U$ enhances the frequency diversity and, hence, makes it possible to satisfy the reliability target $\epsilon_U$ at a larger rate $r_U$. 
%Given $r_U$ and $\Gamma_U$, we use $F_U(r_U,\Gamma_U)$, or only $F_U$, to denote the minimal $F_U$ that satisfies (\ref{URLLCperformance}). 
 
\subsection{mMTC}\label{subsec:mmtc}

The key property of the mMTC traffic is that the set of mMTC devices that transmit in a given radio resource is random and unknown. An mMTC transmission has a fixed rate $r_M$ and consumes one radio resource of $n$ channel uses. Given the rate $r_M$ and the reliability constraint $\epsilon_M$, we focus on the maximum arrival rate $\lambda_M$ that can be supported by the system as the performance criterion of interest. As detailed below, the probability of error measures the fraction of incorrectly decoded devices among the active ones.

%As detailed below, the probability of error is  averaged over all active mMTC devices, similar to~\cite{polyanskiy2017perspective}.

SIC at the BS is a useful strategy to improve the performance of mMTC traffic. As discussed next, a SIC decoder can leverage power imbalances and other mechanisms not reviewed here (see, e.g., \cite{Pratas5g}), in order to sequentially improve the reliability of simultaneous mMTC transmissions. 

To characterize the performance achievable with SIC, we let $[m]$ denote the index of the mMTC device with the $ m$-th largest channel gain $\{G_{m,f_M}\}$ for the allocated frequency $f=f_M$. In the rest of this section, we drop the dependence on $f_M$ for simplicity of notation. By definition, we then have the inequalities $G_{[1]}   \geq G_{[2]}\geq \ldots \geq G_{[{A_M}]}$.
In the absence of interference from eMBB and URLLC traffic, the SINR $\sigma_{[m_0]}$ available when decoding the signal of the $m_0-$th mMTC device, under the additional assumption that the devices with indices $[1],\dots, [m_0-1]$ are correctly decoded, depends only on its channel gain $G_{[m_0]}$ and on the channels gains of the other active mMTC devices as
\begin{equation} \label{SINR_for_mMTC}
\sigma_{[m_0]}=\frac{G_{[m_0]}}{1+\sum_{m =m_0+1}^{A_M} G_{[m]} }.
\end{equation}
The $m_0-$th mMTC device is correctly decoded  if the inequality $\log_2 (1+\sigma_{[m_0]})\geq r_M$ holds; and, if decoding is successful, the signal from the device is subtracted from the received signal. We let  $D_M$ be the random number of mMTC devices in outage, i.e., $D_M$ is the largest integer in $\{0,\ldots,A_M\}$ satisfying, for all $k\in\{1,\ldots, D_M\}$, the inequality
\begin{IEEEeqnarray}{rCl}\label{eq:rate_mMTC}
	\log_2(1+\sigma_{[k]}) \geq r_M
\end{IEEEeqnarray}
The error rate of the mMTC devices is then quantified as the ratio
\begin{equation}
\mathrm{Pr}(E_M)=\frac{\E{D_M}}{\lambda_M}\label{eq:mMTC_error_orth}
\end{equation}between the average number of users in outage, namely $\E{D_M}$, and the average number $\lambda_M$ of active users.
%\kft{Is it better to use the per-user error probability $\frac{1}{\lambda_M}\E{\sum_{m=1}^{A_M} \Pr \left ( \log_2 (1+\sigma_{[m]}) < r_M\right )}$?}
%The probability of error (\ref{eq:mMTC_error_orth}) measures the fraction of mMTC devices that are incorrectly decoded by computing the ratio of the average number of correctly decoded mMTC transmission over the average number of transmissions. 
The maximum rate $\lambda_M$ that can be supported under the reliability condition $\mathrm{Pr}(E_M)=\epsilon_M$ is defined for reference as
\begin{equation}
\lambda_M^{\text{orth}}(r_M)=\textrm{max}\{\lambda_M: \mathrm{Pr}(E_M)\leq \epsilon_M \}\label{eq:mMTCrorth}.
\end{equation}This quantity can be computed by means of Monte Carlo numerical methods.

\section{Slicing for eMBB and URLLC}
\label{sec:eMBBURLLC}

In this section, we consider the coexistence of eMBB and URLLC devices, while assuming that there is no mMTC traffic, i.e., that the mMTC arrival rate is $\lambda_M=0$. 
We first briefly recall, using the results in the previous section, how the performance of the two services can be evaluated for the case of H-OMA, and then analyze the more complex scenario of H-NOMA.

% In order to facilitate the discussion, we define the following events:  
% \begin{itemize}
% \item $\Omega_U$: URLLC transmission is not in outage. 
% \item $\tau_{B,f}$: the eMBB transmission at the $f-$th channel takes place because $G_{B,f} \geq \eta_f$.
% \item $\Omega_{B,f}$: The eMBB transmission at the $f-$th channel is not in outage, which may be caused by the URLLC transmission. 
% \end{itemize}

% Recall that in our model there are no errors caused by noise \gd{Let's remove this. It may confuse a reader not versed with information theory.} and the channel is static within a given radio resource, such that: (1) either all or none of the URLLC transmissions are decoded; (2) either all or none of the eMBB transmissions that took place (for the channels where the event $\tau_{B,f}$ occurs) are decoded.\gd{Is this last statement correct? After all, different eMBB transmit over  frequency radio resources, which fade independently.}

\subsection{Orthogonal Slicing: H-OMA for eMBB and URLLC}

In the case of orthogonal slicing, i.e., under H-OMA, we assume that $F_U$ out of the $F$ frequency radio resources for all minislots in the given radio resource are allocated to the URLLC transmissions, while the remaining $F_B=F-F_U$ radio resources are each allocated to one eMBB user. Note that, in each minislot, the probability that the $F_U$ frequency channels allocated to URLLC traffic are unused is the complement of the activation probability, i.e., $1-a_U$. 

%An example of orthogonal slicing can be constructed using Fig.~\ref{fig:TFmodel}: the mMTC traffic is removed, the URLLC transmissions occupy the first $F_U=4$ channels, and  eMBB traffic is allocated in the remaining $F_B=3$ channels. 
%Note that, in this case, eMBB traffic is not allowed to use the first $F_U=4$ channels, i.e., differently from Fig.~\ref{fig:TFmodel}, channels 2 and 3 cannot be occupied by eMBB traffic.

The performance of the system is specified in terms of the the pair $(r_B,r_U)$ of eMBB sum-rate $r_B$ and URLLC rate $r_U$ achievable at the given reliability levels $(\epsilon_B,\epsilon_U)$. The eMBB sum-rate is obtained as\begin{equation} \label{eq:orth_sum_rB}
r_{B}=(F-F_U)r^{\text{orth}}_B, \qquad \mathrm{[bits/symbol]}
\end{equation}where $r^{\text{orth}}_B$ is obtained as explained in Section \ref{sec:eMBB}. The URLLC rate $r_U$ is computed from (\ref{URLLCperformance}) by imposing the equality $\mathrm{Pr}(E_U)=\epsilon_U$ as detailed in Section \ref{sec:URLLC}.

% The URLLC We have $r_{B,f}=\log_2(1+\beta)$ where $\beta$ is chosen according to $\eta$ for which $\Pr(\Omega_{B,f})=1-q_B(\eta)=\epsilon_B$, such that:

\subsection{Non-orthogonal Slicing: H-NOMA for eMBB and URLLC}\label{subsec:urllc_nonorth}

We now consider non-orthogonal slicing, or H-NOMA, whereby all $F$ frequency channels are used for both eMBB and URLLC transmissions.
Hence, $F_U=F_B=F$. 
With non-orthogonal slicing, eMBB and URLLC transmissions interfere, and, hence, the rate pair $(r_B,r_U)$ cannot be directly obtained from the analysis in Section II.
We next describe different decoding architectures, and derive corresponding achievable pairs $r_B$ and $r_U$ for non-orthogonal slicing.

\emph{Decoding Architectures:} A key observation in the design of decoding schemes is that, due to latency constraints, the decoding of a URLLC transmission cannot wait for, and hence depend upon, the decoding of eMBB traffic. In fact, decoding of a URLLC transmission can only rely on the signal received in the given minislot. This constraint prevents SIC decoders whereby eMBB transmissions are decoded first and canceled from the received signal prior to decoding of the URLLC messages. 
Note also that, because of the heterogeneity of reliability requirements,  decoding eMBB first and then URLLC in a SIC fashion would require decoding the eMBB traffic at the same level of reliability needed for the  URLLC traffic.
As a result of these considerations, in H-NOMA with SIC the URLLC transmissions should be decoded while treating eMBB signals as an additional noise. 

In contrast to URLLC traffic, eMBB requirements are less demanding in terms of latency, and hence eMBB decoding can  wait for URLLC transmissions to be decoded first. This enables a SIC mechanism whereby URLLC messages are decoded and then canceled from the received signal prior to decoding of the eMBB signal. Since the reliability of URLLC is two or more orders of magnitude higher than eMBB, the performance of eMBB under the described SIC decoder is expected to be close to the ideal orthogonal case in which no interference from URLLC traffic is present. This design choice is an instance of reliability diversity. 

That said, the SIC decoder may be ruled out by considerations such as complexity. In such circumstances, one could adopt another decoding approach that is, in a sense, diametrically opposite to SIC in its treatment of URLLC interference. Such a decoder treats any minislot that contains a URLLC transmission as erased or \emph{punctured}. This option of H-NOMA with puncturing is currently being considered within the 5G community~\cite{anand2017joint}. Note that this approach requires the decoder at the BS to be able to detect the presence of URLLC transmissions, e.g., via energy detection. 

\emph{Encoding:} If H-NOMA with SIC is used, we set the eMBB rate to \begin{equation} \label{eq:Maximal_rB_SIC}
r^{\mathrm{SIC}}_{B,f}(G^{\mathrm{tar}}_{B,f})=\log_2(1+G^{\mathrm{tar}}_{B,f}), \qquad \mathrm{[bits/symbol]}
\end{equation}where $G^{\mathrm{tar}}_{B,f}$ represents the target SNR for eMBB transmission, which is to be determined. 

In contrast, in H-NOMA with puncturing and erasure decoding, the eMBB device applies an outer erasure code with rate $1-k/S$, which is concatenated to the codebook used in the physical-layer transmission of the eMBB encoder. Thanks to the erasure code, the decoder is able to correct $k\leq S$ erased minislots, while, if the number of URLLC transmissions is larger than $k$, the decoding process fails. 
The parameter $k$ needs to be designed so as to satisfy the target error rate $\epsilon_B$ for eMBB users. 
The resulting data rate for eMBB transmission in frequency channel $f$ is
\begin{align}\label{eq:ratepunc}
r^{\mathrm{pun}}_{B,f}(G^{\mathrm{tar}}_{B,f},k)=\left(1-\frac{k}{S}\right) \log_2(1+G^{\mathrm{tar}}_{B,f}).
\end{align}

Regarding the selection of the target SNR  $G^{\mathrm{tar}}_{B,f}$, we recall that
in the orthogonal case, as shown in~\eqref{eq:BetaFeta}, the variable $G^{\mathrm{tar}}_{B,f}$ is uniquely determined by the error probability target~$\epsilon_B$ via the threshold SNR $G^{\mathrm{min}}_{B,f}$ defined in~\eqref{eq:Gminexplicit}.
In contrast, with non-orthogonal slicing, it may be beneficial to choose a smaller target SNR  than the one given by~\eqref{eq:BetaFeta}, so as to reduce the interference caused to URLLC transmissions.
This yields the inequality
\begin{IEEEeqnarray}{rCl}
  G^{\mathrm{tar}}_{B,f} \leq \frac{\Gamma_{B}}{ \gamma\mathopen{}\Big(0,\frac{G^{\mathrm{min}}_{B,f}}{\Gamma_{B}}\Big)}. \label{eq:BetaFeta2}
\end{IEEEeqnarray}

%For given $k$ let us use the term ``$k-$punctured eMBB''. 
Summarizing the discussion so far, decoding of URLLC traffic cannot leverage SIC and treats eMBB transmissions as noise. 
In contrast, the eMBB decoder at the BS can either leverage SIC by  decoding URLLC traffic first, or rather treat any minislot occupied by URLLC traffic as erased. 
These are two extreme points among all possible eMBB decoders.

%The case in which the eMBB decoder treats URLLC traffic as noise will be discussed in Section~\ref{sec:DiscussionGeneralSlicing}.

\emph{Rate Region:} The objective of the analysis is to determine the rate region $(r_B,r_U)$ for which the target error probabilities of the two services are satisfied. To this end, we fix the URLLC rate $r_U \in [0,r^{\text{orth}}_U(F)]$, and compute the maximum attainable eMBB rate $r_B$. 
We recall that the available  degrees of freedom in the design are the target SNR $G^{\mathrm{tar}}_{B,f}$ and the minimum channel gain $G^{\mathrm{min}}_{B,f}$ at which an eMBB device is active (or equivalently the activation probability (\ref{eq:actpB})), as well as the erasure code parameter $k$ if a puncturing approach is adopted for eMBB decoding.
We also emphasize that, unlike the orthogonal case, the target SNR $G^{\mathrm{tar}}_{B,f}$ and the minimum SNR $G^{\mathrm{min}}_{B,f}$ are separate degrees of freedom, which are related by the inequality (\ref{eq:BetaFeta2}). 

We start by imposing the reliability constraint for the URLLC user, which yields the following condition for both SIC and erasure decoder:
\begin{equation} \label{r_Udecoding}
\Pr(E_U)=\Pr \lefto( \frac{1}{F_U} \sum_{f=1}^{F_U} \log_2 \lefto(1+\frac{G_{U,f}}{1+\delta_f G^{\mathrm{tar}}_{B,f}} \right) < r_U \right  ) \leq \epsilon_U.
\end{equation}
Here,  $\{\delta_f\}_{f=1}^{F_U}$ are independent Bernoulli random variables with parameter $a_B$ given in (\ref{eq:actpB}).
Recall that $a_B$ is a function of $G^{\mathrm{min}}_{B,f}$. 
The term $\delta_f G^{\mathrm{tar}}_{B,f}$ represents the interference power caused by an eMBB transmission on frequency channel $f$ to the URLLC traffic. The inequality (\ref{r_Udecoding}) imposes a joint constraint on both  $G^{\mathrm{tar}}_{B,f}$ and $G^{\mathrm{min}}_{B,f}$. Next, we impose the reliability constraint for eMBB traffic by considering separately SIC and erasure decoders.

% Given $r_U, \Gamma_U, \Gamma_B,$ and $F$, the objective is to choose the activation probability $a_B$ and $G^{\mathrm{tar}}_{B}$ such that \eqref{r_Udecoding} is satisfied with equality. We note that the choice of the threshold $G^{\mathrm{min}}_{B}$ uniquely determines $a_B$, but not $G^{\mathrm{tar}}_{B}$, since the latter has an additional degree of freedom due to (\ref{eq:BetaFeta2}). 

\subsubsection{SIC decoder} 
Under H-NOMA the decoding of an eMBB message is generally affected by the interference from the URLLC users. 
However, this interference is not present if: (\emph{i}) there are no URLLC transmissions, i.e., $S_U=0$; or (\emph{ii}) if URLLC transmissions are present, i.e., $S_U>0$, but the corresponding signals are decoded successfully and canceled by the SIC decoder.
 As for the latter event, since the interference from eMBB users and  the fading gains are constant across the minislots, either all URLLC transmissions are decoded incorrectly (event $E_U$) or they are all correctly decoded (event $\overline{E}_U$). 

Based on the discussion above, we can bound the eMBB error probability by distinguishing the case in which the eMBB transmission is subject to interference from URLLC signals, and the case in which is not, using the law of total probability, as follows:
\begin{IEEEeqnarray}{rCl}
\Pr(E_B)&=&\Pr(S_U=0)\Pr(E_B|S_U=0)\nonumber \\
&&{}+\Pr(S_U>0)\big(\Pr(E_U|S_U>0)\Pr(E_B|E_U,S_U>0)\nonumber\\
&&\qquad\qquad\qquad\quad{}+\Pr(\bar{E}_U|S_U>0)\Pr(E_B|\bar{E}_U,S_U>0)\big)  \label{eq:PrEB_cond_1}\\
&{=}&(1-a_U)^S(1-a_B)+[1-(1-a_U)^S]\big(\epsilon_U \Pr(E_B|E_U,S_U>0)\nonumber\\
&&\qquad\qquad\qquad\qquad\qquad\qquad\qquad\qquad{}+(1-\epsilon_U)\Pr(E_B|\bar{E}_U,S_U>0)\big)\IEEEeqnarraynumspace \label{eq:PrEB_cond_2}\\
&{\leq}&(1-a_U)^S(1-a_B)+[1-(1-a_U)^S]\left(\epsilon_U +(1-\epsilon_U)(1-a_B) \right). \label{eq:PrEB_cond}
\end{IEEEeqnarray}
%
% This is presented through (\ref{eq:PrEB_cond}). In this equation,
% \begin{figure*}[!t]
% % ensure that we have normalsize text
% \normalsize
% % Store the current equation number.
% %\setcounter{MYtempeqncnt}{\value{equation}}
% % Set the equation number to one less than the one
% % desired for the first equation here.
% % The value here will have to changed if equations
% % are added or removed prior to the place these
% % equations are referenced in the main text.
% %\setcounter{equation}{5}
% \begin{IEEEeqnarray}{rCl}
% \Pr(E_B)&=&\Pr(S_U=0)\Pr(E_B|S_U=0)\nonumber \\
% &&{}+\Pr(S_U>0)\big(\Pr(E_U|S_U>0)\Pr(E_B|E_U,S_U>0)\nonumber\\
% &&\qquad\qquad\qquad\quad{}+\Pr(\bar{E}_U|S_U>0)\Pr(E_B|\bar{E}_U,S_U>0)\big)  \\
% &\stackrel{\mathrm{(a)}}{=}&(1-a_U)^S(1-a_B)+[1-(1-a_U)^S]\big(\epsilon_U \Pr(E_B|E_U,S_U>0)\nonumber\\
% &&\qquad\qquad\qquad\qquad\qquad\qquad\qquad\qquad{}+(1-\epsilon_U)\Pr(E_B|\bar{E}_U,S_U>0)\big)\IEEEeqnarraynumspace \\
% &\stackrel{\mathrm{(b)}}{\leq}&(1-a_U)^S(1-a_B)+[1-(1-a_U)^S]\left(\epsilon_U +(1-\epsilon_U)(1-a_B) \right)  \label{eq:PrEB_cond}
% \end{IEEEeqnarray}
% % Restore the current equation number.
% %\setcounter{equation}{\value{MYtempeqncnt}}
% % IEEE uses as a separator
% \hrulefill
% % The spacer can be tweaked to stop underfull vboxes.
% \vspace*{4pt}
% \end{figure*}
Here, equality~\eqref{eq:PrEB_cond_1} holds because the only source of outage for eMBB in absence of URLLC interference is the instantaneous SNR being below the minimum SNR, which implies $\Pr(E_B|S_U=0)=1-a_B$; moreover,~\eqref{eq:PrEB_cond_2} follows by using that $\Pr(E_B|E_U,S_U>0) \leq 1$ and that canceling URLLC interference results in the same performance achievable in the absence of URLLC transmissions, so that we have the equality $\Pr(E_B|\bar{E}_U,S_U>0)=\Pr(E_B|S_U=0)=1-a_B$.
Imposing the reliability condition $\Pr(E_B)\leq \epsilon_B$ and using~\eqref{eq:PrEB_cond} we obtain the inequality
\begin{align}\label{eq:aBinURLLCeMBB}
a_B \geq \frac{1-\epsilon_B}{1-\epsilon_U(1-(1-a_U)^S)}.
\end{align}
As already pointed out, this equivalently imposes a constraint on $G^{\mathrm{min}}_{B,f}$ through (\ref{eq:actpB}). 

From (\ref{eq:aBinURLLCeMBB}), we see that, unlike the orthogonal case, with  non-orthogonal slicing the eMBB activation probability $a_B$ is larger than $1-\epsilon_B$. This is becasue URLLC interference may cause an eMBB decoding error even when the eMBB's SNR is above the threshold. However, the impact of URLLC interference is typically minimal.
Indeed, the high reliability requirements for URLLC, which are reflected by the very small value of $\epsilon_U$, imply that $a_B$ is close to $1-\epsilon_B$.

To summarize, for a given feasible URLLC rate $r_U \in [0,r^{\text{orth}}_U(F)]$, the maximum eMBB rate is obtained by maximizing $\log_2(1+G_{B,f}^{\mathrm{tar}})$ subject on the constraints on $G_{B}^{\mathrm{tar}}$ and $G^{\mathrm{min}}_{B}$ implied by (\ref{eq:BetaFeta2}), (\ref{r_Udecoding}), and (\ref{eq:aBinURLLCeMBB}).
This maximization requires the use of a two-dimensional numerical search. 
Note that the activation probability $a_B$ is typically very close to $1$, and hence, when solving this problem, one can conservatively assume that the eMBB interference is always present in (\ref{r_Udecoding}), i.e., $\delta_f=1$. 
In contrast, the dependence of the right-hand side of~\eqref{r_Udecoding} on $G_{B}^{\mathrm{tar}}$ causes a non-trivial interdependence between $r_U$ and $r_B$.

\subsubsection{Puncturing and erasure decoder} 
Turning now to the erasure decoder, we can write the probability of error for an eMBB user  by means of the law of total probability as
\begin{IEEEeqnarray}{rCl}\label{eq:Eb_S=sk_1}
\Pr(E_B)&=& \Pr(S_U \leq k) \Pr(E_B|S_U \leq k)+\Pr(S_U > k) \Pr(E_B|S_U > k),
\end{IEEEeqnarray}
where we have distinguished the case in which the erasure code is able to correct the erasures caused by URLLC transmissions, i.e., $S_U \leq k$, and the case in which an error is instead declared, i.e., $S_U > k$. 
When the latter event occurs, a decoding error occurs, and hence we have $\Pr(E_B|S_U > k)=1$. 
In contrast, when $S_U \leq k$, the only source of outage is the instantaneous SNR being below threshold, which results in $\Pr(E_B|S_U \leq k)=1-a_B$. 
Overall, the resulting eMBB reliability requirement is
\begin{align}\label{eq:Eb_S=sk}
\Pr(E_B)&= \Pr(S_U \leq k)(1-a_B)+\Pr(S_U > k)\leq\epsilon_B.
\end{align}

Imposing equality in (\ref{eq:Eb_S=sk}), we determine the parameter $a_B$ and, hence, $G^{\mathrm{min}}_{B,f}$ via~\eqref{eq:actpB}. 
Given the desired feasible URLLC rate $r_U \in [0,r^{\text{orth}}_U(F)]$, , we then obtain the target SNR $G^{\mathrm{tar}}_{B,f}$ and, hence, the eMBB rate (\ref{eq:ratepunc}) from the URLLC reliability condition (\ref{r_Udecoding}).

\subsection{Numerical Illustration}

% In order to get an insight into the relation between $r_B, r_U, \epsilon_B,\epsilon_U$, we apply tail bounding technique in (\ref{r_Udecoding}). Note that the orthogonal case is covered by setting $\Pr(\delta_f=0)=1$, while the non-orthogonal case is covered by setting $F_U=F$ and $\Pr(\delta_f=0)=\epsilon_B$.\gd{Is this so? What about (18)? The relation between activation and error is more complicated in the non-orthogonal case, right?} Appendix~\ref{sec:LBr_U} presents the derivation of lower bounds on $r_U$. When orthogonal allocation is used we put $G^{\mathrm{tar}}_{B}=0$ in (\ref{eq:r_U_Delta_t}), leading to:\gd{This forward reference is confusing. I think the flow needs to be revised here.}
% \begin{align}\label{eq:r_U_LB_orth}
% r_U=\frac{1}{tF_U}\log_2 \epsilon_U + \log_2 \left(1+\Gamma_U \right) - \frac{1}{t} \log_2 \Delta_t 
% \end{align}
% while when SIC is used we set $F=F_U$ and (\ref{eq:r_U_Delta_t}) can be written as:
% \begin{align} \label{eq:r_U_LB_SIC}
% r_U=\frac{1}{tF}\log_2 \epsilon_U + \log_2 \left(1+\frac{\Gamma_U}{1+G^{\mathrm{tar}}_{B}} \right) - \frac{1}{t} \log_2 \Delta_t 
% \end{align}

\begin{figure}
\centering
\includegraphics[width=8.3cm]{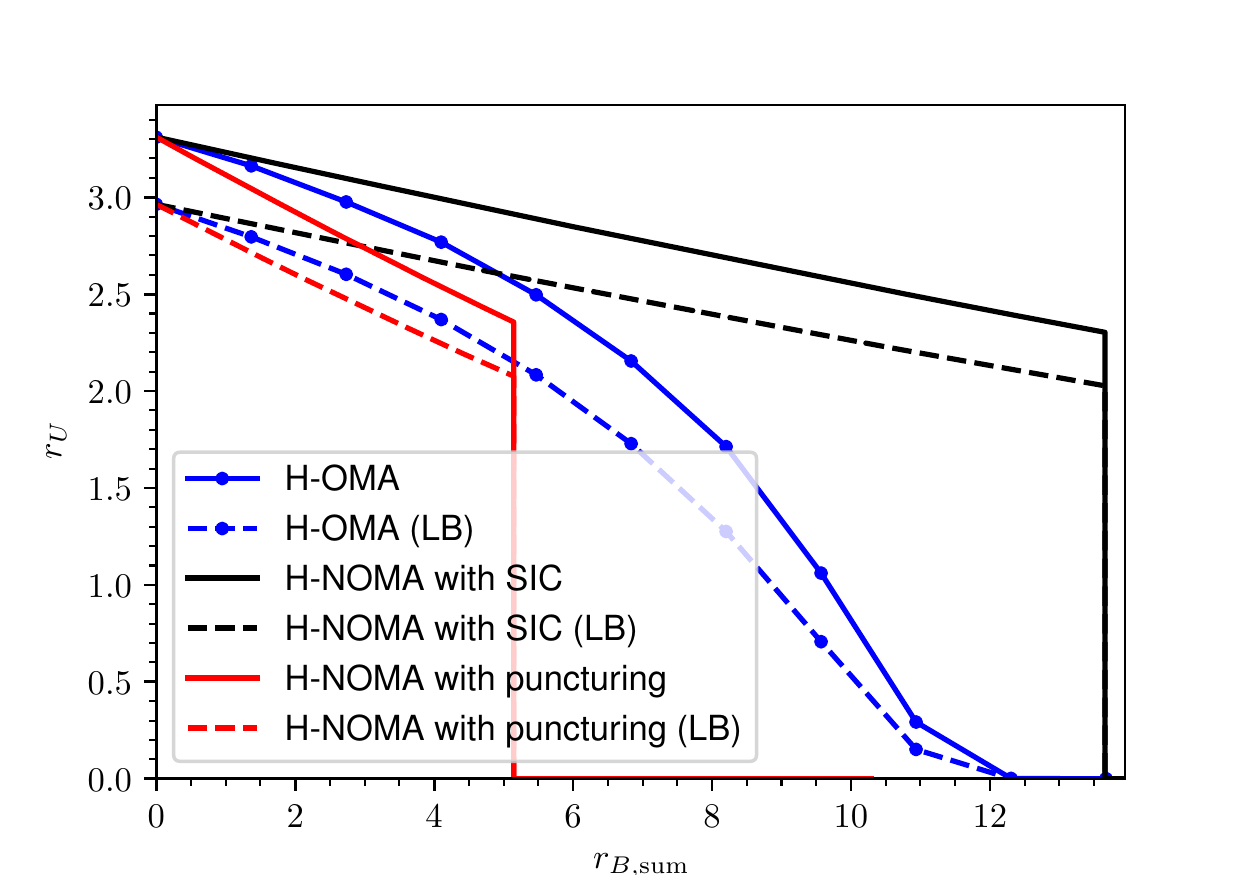}
  \caption{Rate region $(r_{B,\text{sum}},r_U)$ for the eMBB rate $r_{B,\text{sum}}$ and the URLLC rate $r_U$ when $\Gamma_U=20$ dB, $\Gamma_B=10$ dB, $S=5, a_U=0.1, F=10, \epsilon_U=10^{-5}, \epsilon_B=10^{-3}$. H-NOMA is present with two variants, SIC and puncturing. The lower bound (LB) is derived in Appendix A.}
 \label{fig:embb_urllc_GamUdB_20_GamBdB_10_S_5_aU_0_10_F_10_epsU_5}
\end{figure}
\begin{figure}
\centering
  \includegraphics[width=8.3cm]{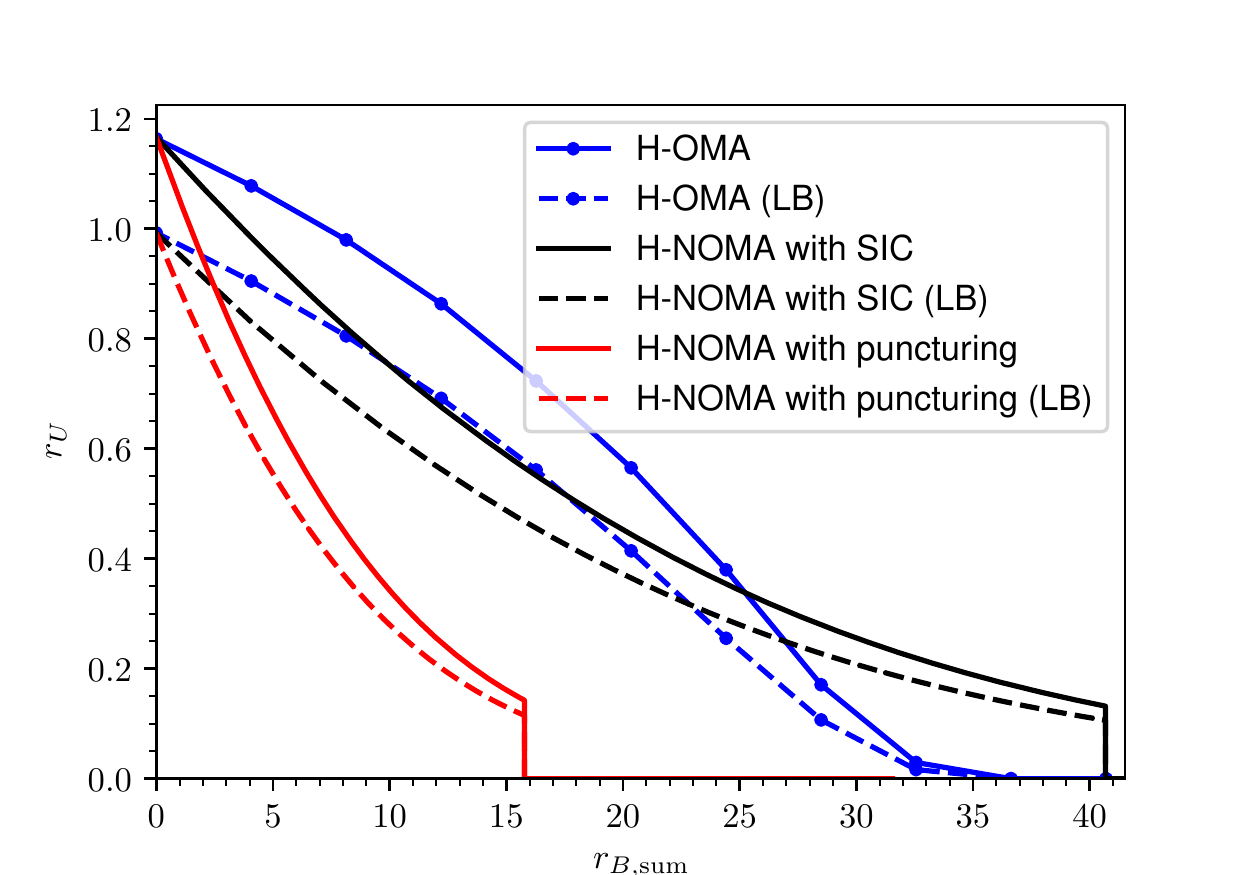}
  \caption{Rate region $(r_{B,\text{sum}},r_U)$ for the eMBB rate $r_{B,\text{sum}}$ and the URLLC rate $r_U$ when $\Gamma_U=10$ dB, $\Gamma_B=20$ dB, $S=5, a_U=0.1, F=10, \epsilon_U=10^{-5}, \epsilon_B=10^{-3}$. H-NOMA is present with two variants, SIC and puncturing. The lower bound (LB) is derived in Appendix A.}
 \label{fig:embb_urllc_GamUdB_10_GamBdB_20_S_5_aU_0_10_F_10_epsU_5}
\end{figure}
% Here $\Delta_t \geq 1$ is a convexity factor\gd{??} and $t>0$ is a parameter that can be chosen to maximize the lower bound. 
% The following can be observed: (1) In both orthogonal and SIC decoder, $r_U$ decreases logarithmically with the decrease of the required error rate $\epsilon_U$. (2) the increase of $r_B$ increases $G^{\mathrm{tar}}_{B}$ and hence decreases $r_U$; (3) when $\epsilon_B$ is small, which means $a_B$ is close to $1$, it does not affect $r_U$ significantly in the case of non-orthogonal slicing, i.e. we do not lose much in choosing $r_U$ by assuming that eMBB is always active on all $F$ channels. 

Here we present simulation results for the rate region $(r_{B},r_U)$ for H-OMA as well as H-NOMA, with both SIC and puncturing decoders. 
In addition to the results obtained from the previous analysis, we also show curves obtained from the expressions derived in Appendix~\ref{sec:LBr_U}, which are easier to evaluate and are shown to provide a performance lower bound (``LB"). 

In Figs.~\ref{fig:embb_urllc_GamUdB_20_GamBdB_10_S_5_aU_0_10_F_10_epsU_5}  and  \ref{fig:embb_urllc_GamUdB_10_GamBdB_20_S_5_aU_0_10_F_10_epsU_5}, we plot the rate regions for $S=5, a_U=0.1, F=10, \epsilon_U=10^{-5}, \epsilon_B=10^{-3}$. Fig.~\ref{fig:embb_urllc_GamUdB_20_GamBdB_10_S_5_aU_0_10_F_10_epsU_5} considers the case $\Gamma_U>\Gamma_B$ with $\Gamma_U=20$ dB and $\Gamma_B=10$ dB, while Fig. \ref{fig:embb_urllc_GamUdB_10_GamBdB_20_S_5_aU_0_10_F_10_epsU_5} focuses on the complementary set-up with $\Gamma_B>\Gamma_U$ when $\Gamma_U=10$ dB and $\Gamma_B=20$ dB. For both figures, H-NOMA with puncturing uses the optimal puncturing parameter $k$. 

% \color{green} I guess that "LB" refers to the simplified analysis? Why is it only for SIC? What do we gain from looking at this curve?\color{black} 
 
%The Monte Carlo evaluation of (\ref{eq:r_U_CorrectEt}) is much more efficient than the accurate calculation based on (\ref{eq:BetaFeta2}), (\ref{r_Udecoding}) and (\ref{eq:aBinURLLCeMBB}), as described above. The lower bound (\ref{eq:r_U_CorrectEt}) can be used for all three cases: (i) Setting $G^{\mathrm{tar}}_{B}=0$ covers teh orthogonal slicing; (ii) Setting $F_U=F$ covers SIC and puncturing, as the behavior of $r_U$ in both cases is identical. The lower bound (\ref{eq:r_U_CorrectEt}) provides an important insight: $r_U$ decreases logarithmically as the URLLC error rate $\epsilon_U$ decreases, i.e. the reliability becomes more stringent.  

For both set-ups considered in the figure, H-MONA with puncturing is outperformed by both H-OMA and H-NOMA with SIC. Furthermore, when $\Gamma_U>\Gamma_B$, we see from Fig.~\ref{fig:embb_urllc_GamUdB_20_GamBdB_10_S_5_aU_0_10_F_10_epsU_5} that the SIC region dominates the region achievable by orthogonal slicing. This is thanks to the capability of the BS to decode and cancel URLLC transmissions by leveraging reliability diversity.

In contrast, when $\Gamma_U<\Gamma_B$, Fig.~\ref{fig:embb_urllc_GamUdB_10_GamBdB_20_S_5_aU_0_10_F_10_epsU_5} shows that orthogonal slicing can attain pairs $(r_{B},r_U)$ that are not attainable by H-NOMA with SIC. In particular, H-OMA is preferable if one wishes to obtain large values of the URLLC rate. This is due to the difficulty of ensuring high reliability in the presence of eMBB transmissions when $\Gamma_U<\Gamma_B$. We recall that this is a consequence of the impossibility to decode and cancel eMBB transmissions prior to URLLC decoding owing to the URLLC latency constraint. In contrast, if one is interested in guaranteeing large eMBB sum-rates, H-NOMA offers significant performance gains. This is because non-orthogonal transmission allows eMBB users to operate over a larger number of spectral resources while not being significantly affected by URLLC interference.

% Finally, in both Figs.~\ref{fig:embb_urllc_GamUdB_20_GamBdB_10_S_5_aU_0_10_F_10_epsU_5}  and \ref{fig:embb_urllc_GamUdB_10_GamBdB_20_S_5_aU_0_10_F_10_epsU_5} we have plotted the regions obtained by lower-bounding $r_U$ using the expression we provide in Appendix~\ref{sec:LBr_U}, eq.~\eqref{eq:r_U_CorrectEt}. 

We see that the lower bound is able to capture the shape of the region obtained through more accurate and time-consuming Monte-Carlo simulations.

\section{Slicing for eMBB and mMTC}
\label{sec:eMBBmMTC}

In this section, we treat the slicing of wireless resources to jointly support eMBB and mMTC services, while assuming that the URLLC traffic, if present, has been allocated orthogonal resources. Analogously to the case of eMBB-URLLC coexistence, we consider separately orthogonal slicing (H-OMA) and non-orthogonal slicing (H-NOMA). 
We shall focus without loss of generality on the case $F=1$, since the mMTC users are assumed to be active on a single frequency channel. The extension to the case $F>1$, in which the mMTC devices are allowed to randomly access all $F$ channels, is rather straightforward, as further elaborated in Section~\ref{sec:extensions}. Since a single channel is considered, in this section we omit all frequency indices $f$.

\subsection{Orthogonal Slicing: H-OMA for eMBB and mMTC}
For the case of orthogonal slicing, we assume that the eMBB and the mMTC devices use the  frequency radio resource in a time-sharing manner. Let $\alpha\in[0,1]$ and $1-\alpha$ be fraction of time in which the resources are allocated to the eMBB device and the mMTC devices, respectively. We aim at characterizing the region of pairs $(r_B, \lambda_{M})$ of eMBB rate $r_B$ and mMTC arrival rate $\lambda_{M}$ that can be supported by orthogonal slicing for a given mMTC transmission rate requirement $r_M$ and probability of error $\epsilon_M$.

For a given time-sharing factor $\alpha$, the achievable pair of eMBB rate $r_B$ and mMTC arrival rate $\lambda_M$ can be written in terms of the quantities derived in Sections~\ref{sec:eMBB} and \ref{subsec:mmtc} as
\begin{IEEEeqnarray}{rCl}
r_B &=& \alpha r^{\text{orth}}_B \\ 
\lambda_M &=& \lambda_M^{\text{orth}}\mathopen{}\left(\frac{r_M}{1-\alpha}\right),\label{eq:OrthogonalSlicingmMTC}
\end{IEEEeqnarray}respectively, where $r^{\text{orth}}_B$ is obtained as explained in Section \ref{sec:eMBB} and $\lambda_{M}^{\text{orth}}(\cdot)$ is defined in (\ref{eq:mMTCrorth}). In fact, with orthogonal slicing, both the achievable eMBB rate $r_B$ and the achievable mMTC transmission rate (specified on the right-hand-side of (\ref{eq:rate_mMTC})) are scaled according to the fraction of time resources allocated to the service. 

\subsection{Non-orthogonal Slicing: H-NOMA for eMBB and mMTC}

In H-NOMA, the eMBB device is allowed to use the radio resource at the same time as the mMTC devices. 
%We discuss interference management mechanisms at the BS.

%learly, the region $(r_B, \lambda_{M}^{\text{non-orth}})$ will be changed compared to the orthogonal slicing. On the other hand, some of the points from the region are easy to be seen. For example, when $\lambda_{M}^{\text{non-orth}} \approx 0$, the value of $r_B$ will approach the value attainable in orthogonal slicing with $\alpha \approx 1$. 

\emph{Decoding Achitecture}
As argued in Section~\ref{subsec:mmtc}, a SIC decoder may enhance the reliability of mMTC decoding. Furthermore, when radio resources are allocated exclusively to mMTC devices, optimal decoding follows the order of descending channel gains. The situation is more complicated in the presence of an interfering eMBB transmission.

In light of the higher reliability requirements of eMBB transmissions as compared to  mMTC traffic, i.e., $\epsilon_B \ll \epsilon_M$, one may be tempted to consider decoding the eMBB traffic before attempting to decode any mMTC traffic. 
This appears to be in line with the discussion in the previous section concerning SIC for eMBB and URLLC coexistence. 
However, this approach is suboptimal, since it neglects to account for the different definition of reliability of mMTC traffic. In fact, the probability of error (\ref{eq:mMTC_error_orth}) measures the fraction of incorrectly detected active users and not a per-device decoding probability. As such, some of the active mMTC devices may well have very high channel gains, hence, causing large interference, making it beneficial to decode and cancel them prior to decoding the eMBB signal. Selecting a SIC decoder that accounts for this important feature of mMTC traffic is another example of a design choice that utilizes reliability diversity.

Based on this discussion, we assume that, at each decoding step, the BS decodes  either the eMBB device, provided that it has not been decoded yet, or the next available mMTC device in order of decreasing channel gains. Note that this implies that the decoding step at which the eMBB device is decoded is random, as it depends on the realization of the channel gains. The process ends when no more transmissions can be reliably decoded. 
%Specifically, the decoding procedure takes up to $A_M+1$ rounds. In the beginning of the $\ell$-th round,  $D_M\in\{0,\ldots,A_M\}$ and $D_B\in\{0,1\}$ denotes the number of decoded mMTC and eMBB devices at decoded so far. 

As in non-orthogonal slicing of eMBB and URLLC, the eMBB rate is set to
\begin{IEEEeqnarray}{rCl}
 r_B &=&  \log_2(1+G_B^{\text{tar}})
\end{IEEEeqnarray}
where $G_B^{\text{tar}}$ is the target SNR for the eMBB transmission, which is to be determined. Similar to the eMMB-URLLC coexistence case (see Section~\ref{subsec:urllc_nonorth}), this quantity needs to satisfy
\begin{IEEEeqnarray}{rCl}
	  G^{\mathrm{tar}}_{B} \leq \frac{\Gamma_{B}}{ \gamma\mathopen{}\Big(0,\frac{G^{\mathrm{min}}_{B}}{\Gamma_{B}}\Big)}. \label{eq:BetaFeta3}
\end{IEEEeqnarray}
Again, as in Section~\ref{subsec:urllc_nonorth}, we allow the eMBB device not to use the maximal power since it may be beneficial to use a value of $G_B^\text{tar}$ lower than the right-hand side of \eqref{eq:BetaFeta3} in order to control the impact of eMBB interference on the overall SIC procedure.  

Next, we formalize the SIC decoding procedure. When the eMBB is inactive because of an insufficient SNR, i.e., $G_B < G_B^{\text{min}}$, the SIC decoding procedure is equivalent to the procedure described in Section~\ref{subsec:mmtc}, namely, the mMTC devices are decoded in the order of decreasing channel gains. 
When the eMBB is active, the SIC procedure runs as follows. Starting from $m_0=1$, the receiver computes the SINR for the $m_0$-th mMTC device as 
\begin{IEEEeqnarray}{rCl}
	\sigma_{[m_0]} &=& \frac{G_{[m_0]}}{1 + G_B^{\text{tar}} + \sum_{m=m_0 + 1}^{A_M} G_{[m]} }.
\end{IEEEeqnarray}
If $\log_2(1+\sigma_{[m_0]} ) \geq r_M$, the $m_0$-th mMTC is decoded, canceled, $m_0$ is incremented by one, and the procedure starts over. Otherwise, the receiver attempts to decode the eMBB user. To this end, it computes the SINR of the eMBB transmission as
\begin{IEEEeqnarray}{rCl}
	\sigma_{B} &=& \frac{G_B^{\text{tar}}}{1 + \sum_{m=m_0}^{A_M} G_{[m]} }
\end{IEEEeqnarray}
and decodes and cancels the eMBB if the condition $\log_2(1+G_B^{\text{tar}}) \geq r_B$ is satisfied. If the eMBB is decoded successfully, the decoding procedure continues as in Section~\ref{subsec:mmtc}. If $\log_2(1+G_B^{\text{tar}})< r_B$, the procedure terminates. 

Let $D_M\in\{0,\ldots,A_M\}$ and $D_B\in\{0,1\}$ be the random variables denoting the number of decoded mMTC and eMBB devices. With this notation, the probabilities of error for mMTC and eMBB users are given as $\mathrm{Pr}(E_M) = 1-\E{D_M}/\lambda_M$ and $\mathrm{Pr}(E_B) = 1-\E{D_B}$, respectively. 

In order to characterize the   achievable pairs ($r_B$, $\lambda_M$), we evaluate the maximum supported mMTC arrival rate $\lambda_{M}$ as a function the eMBB rate $r_B$ as
\begin{IEEEeqnarray}{rCl} 
\lambda_{M}^{\text{non-orth}}(r_{B})&=& \max\Big\{\lambda_M\geq 0: \exists G^{\text{tar}}_{B} \text{ and } G_{B}^{\text{min}} \text{ s.t. } \E{D_M} / \lambda_M \geq 1-\epsilon_M \nonumber\\
 && \qquad\qquad\qquad\qquad\qquad\qquad\qquad\  \    \text{ and } \E{D_B} \geq 1-\epsilon_B \Big\}.\label{eq:Rdef}
\end{IEEEeqnarray}
We remark that, the probability distributions of $D_M$ and $D_B$ depend on the parameters $\lambda_M$, $G^{\text{tar}}_{B}$, $G_{B}^{\text{min}}$, $r_{B}$, and $r_M$. 
The computation of \eqref{eq:Rdef} requires Monte Carlo simulations. % for each set of the parameters in order to extract the values of the borders of the region ($r_B$,$\lambda_M$).

\subsection{Numerical Illustration}

We present numerical simulation results illustrating the  trade-offs between the eMBB rate $r_B$ and the mMTC arrival rate $\lambda_M$ for orthogonal and non-orthogonal slicing. In addition to numerical results obtained by solving~\eqref{eq:Rdef} through Monte Carlo methods, we also report results obtained upper and lower bounds on $\lambda_M^{\text{non-orth}}(\cdot)$ in (\ref{eq:Rdef}), which are easier to evaluate and are derived in  Appendix~\ref{sec:AppeMBBmMTC}. Throughout this section, we set $\epsilon_M=10^{-1}$, and $r_M = 0.04$. 

\begin{figure} 
\centering
  \includegraphics[width=8.3cm]{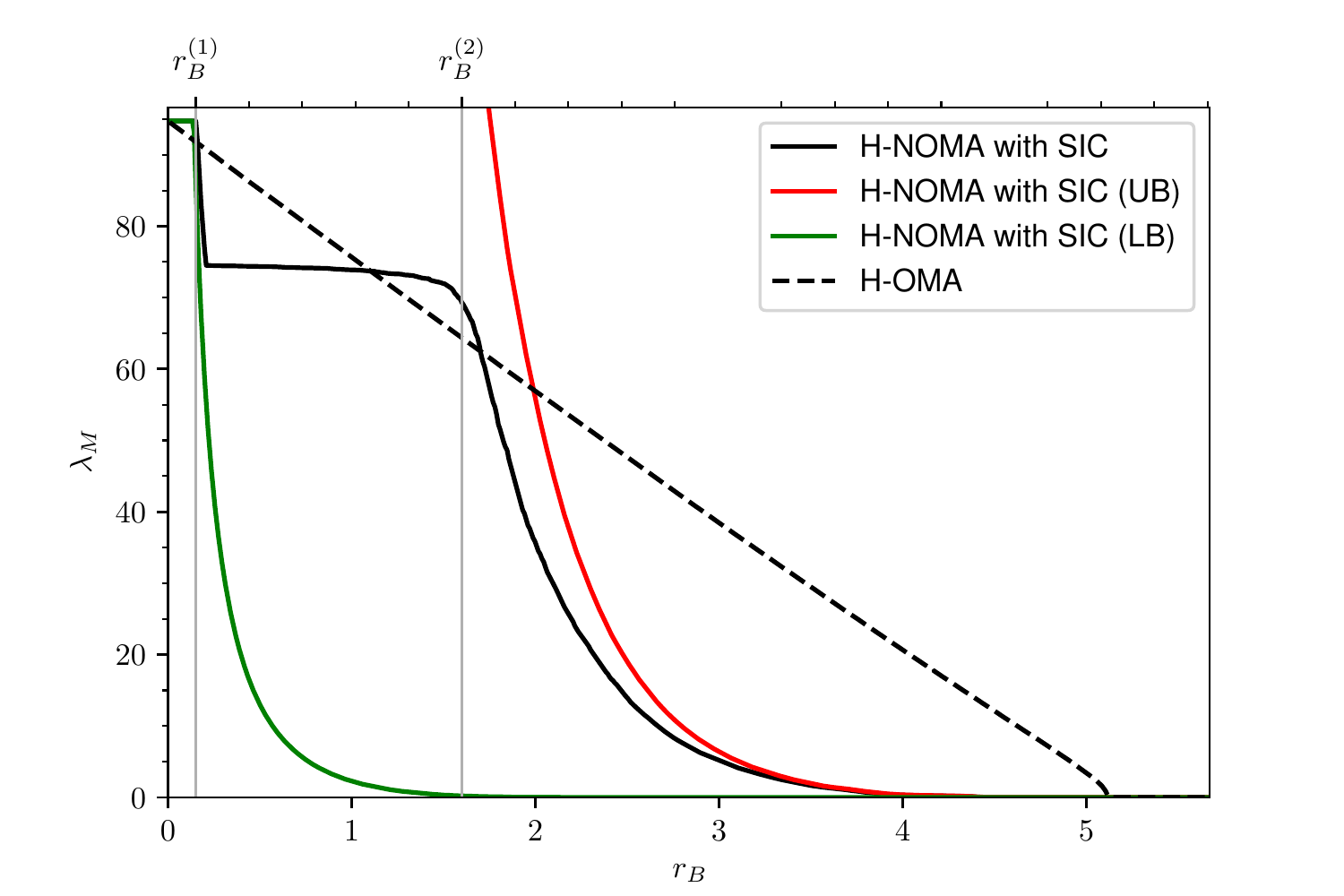}
  \caption{Arrival rates $\lambda_M^{\text{orth}}$ and $\lambda_M^{\text{non-orth}}$ for mMTC traffic under H-OMA and H-NOMA, respectively, as a function of the eMBB rate $r_B$. The upper bounds (UB) and lower bounds (LB) on the H-NOMA arrival rate $\lambda_M^{\text{non-orth}}(r_B)$ derived in Appendix B are also shown. The parameters are $\Gamma_M=5$ dB, $\Gamma_B=25$ dB, $\epsilon_M=10^{-1}, \epsilon_B=10^{-3}$, and $r_M = 0.04$. }
 \label{fig:embb_mmtc_fig1_GamMdB_5_GamBdB_25_rM_0_0400_epsM_1_epsB_3}
\end{figure}

In Fig.~\ref{fig:embb_mmtc_fig1_GamMdB_5_GamBdB_25_rM_0_0400_epsM_1_epsB_3}, we plot the maximum mMTC arrival rate $\lambda_M$ for both orthogonal and non-orthogonal slicing as a function of $r_B$ when $\Gamma_M=5$ dB, $\Gamma_B=25$ dB, and $\epsilon_B=10^{-3}$.
%We have also plotted the average number of mMTC devices decoded before the eMBB device is decoded, which is very indicative about the changes in the operating regime. 
When orthogonal slicing (H-OMA) is used, the supported mMTC arrival rate $\lambda_M$ is seen to decrease in an approximately linear fashion with the eMBB rate. As for non-orthogonal slicing (H-NOMA), we observe three fundamentally different regimes as $r_B$ changes from zero towards is maximal value $r_B^{\text{orth}}$.

%Note that it is not a perfect straight line between the points $(0,\lambda_M^{\text{orth}}(0))$ and $(r_{B,\text{max}},0)$ because $\lambda_M^{\text{orth}}(0)$ is not inversely proportional with $r_M$.

%\PP{In order to understand these regimes, Fig.~\ref{fig:embb_mmtc_fig1_GamMdB_5_GamBdB_25_rM_0_0400_epsM_1_epsB_3} also contains the curve that depicts the average number of mMTCs that \emph{can} be decoded before the eMBB device is decoded. Note that this is different from the curve that would depict how many mMTC devices \emph{must} be decoded before the eMBB device is decoded.}

The \emph{first regime} consists of very small values of the eMMB rate $r_B$, for which the supported arrival rate $\lambda_M^{\text{non-orth}}$ is almost constant. At such values of $r_B$, the eMBB device can be reliably decoded before the mMTC devices. Therefore, interference from eMBB user can be cancelled, and the performance of mMTC traffic is unaffected by small increases in $r_B$. The \emph{second regime} spans intermediate values of $r_B$. In this case, the eMBB signal can only be decoded after some of the strongest mMTC signals are decoded and canceled. Hence, the mMTC performance is reduced by the interference from eMBB transmissions. Also, the SIC decoder tends to stop the decoding process after detecting the eMBB user, and decoding typically fails while detecting an mMTC device because of the interference from the other, yet undecoded, mMTC devices. 
In the \emph{third regime}, eMBB decoding  fails with a probability comparable to that of the weaker mMTC devices due to the mutual interference between the two services. 
As a result, in this regime, the supported mMTC arrival rate decays to zero as the eMBB rate $r_B$ increases. 

The first and the third regime identified in Fig.~\ref{fig:embb_mmtc_fig1_GamMdB_5_GamBdB_25_rM_0_0400_epsM_1_epsB_3} can also be understood with the help of the lower and upper bounds derived in Appendix~\ref{sec:AppeMBBmMTC}. In particular, when $r_B$ is very low, as mentioned, it is almost always possible to decode the eMBB transmission before decoding any mMTC device. 
This is the premise of the lower bound, which, as shown in  Fig.~\ref{fig:embb_mmtc_fig1_GamMdB_5_GamBdB_25_rM_0_0400_epsM_1_epsB_3}, agrees with the simulation results in the first regime. 
On the contrary, the upper bound is computed by first identifying the subset of mMTC devices whose channels are so weak that the additive noise and the eMBB interference alone make their decoding impossible. 
The upper bound is seen to agree the simulation results in the third regime, i.e., when $r_B$ is large. 

\begin{figure} 
\centering
  \includegraphics[width=8.3cm]{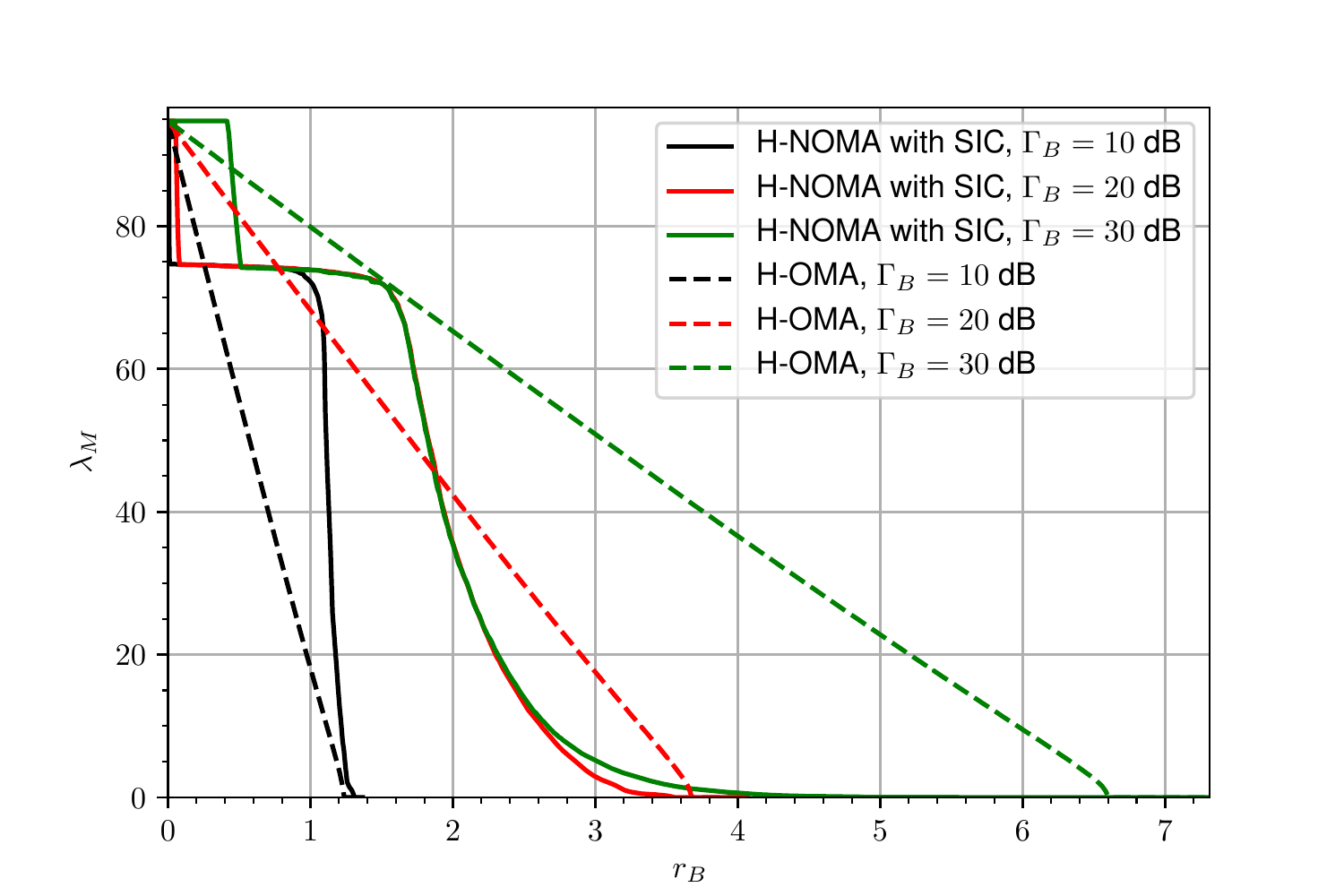}
  \caption{Arrival rates $\lambda_M^{\text{orth}}$ and $\lambda_M^{\text{non-orth}}$ for mMTC traffic under H-OMA and H-NOMA, respectively, as a function of the eMBB rate $r_B$ for $\Gamma_B \in \{10, 20, 30\}$ dB. The parameters are $\Gamma_M=5$ dB, $\epsilon_M=10^{-1}, \epsilon_B=10^{-3}$, and $r_M = 0.04$.}
 \label{fig:embb_mmtc_fig2_GamMdB_5_rM_0_0400_epsM_1_epsB_3}
\end{figure}

\begin{figure} 
\centering
  \includegraphics[width=8.3cm]{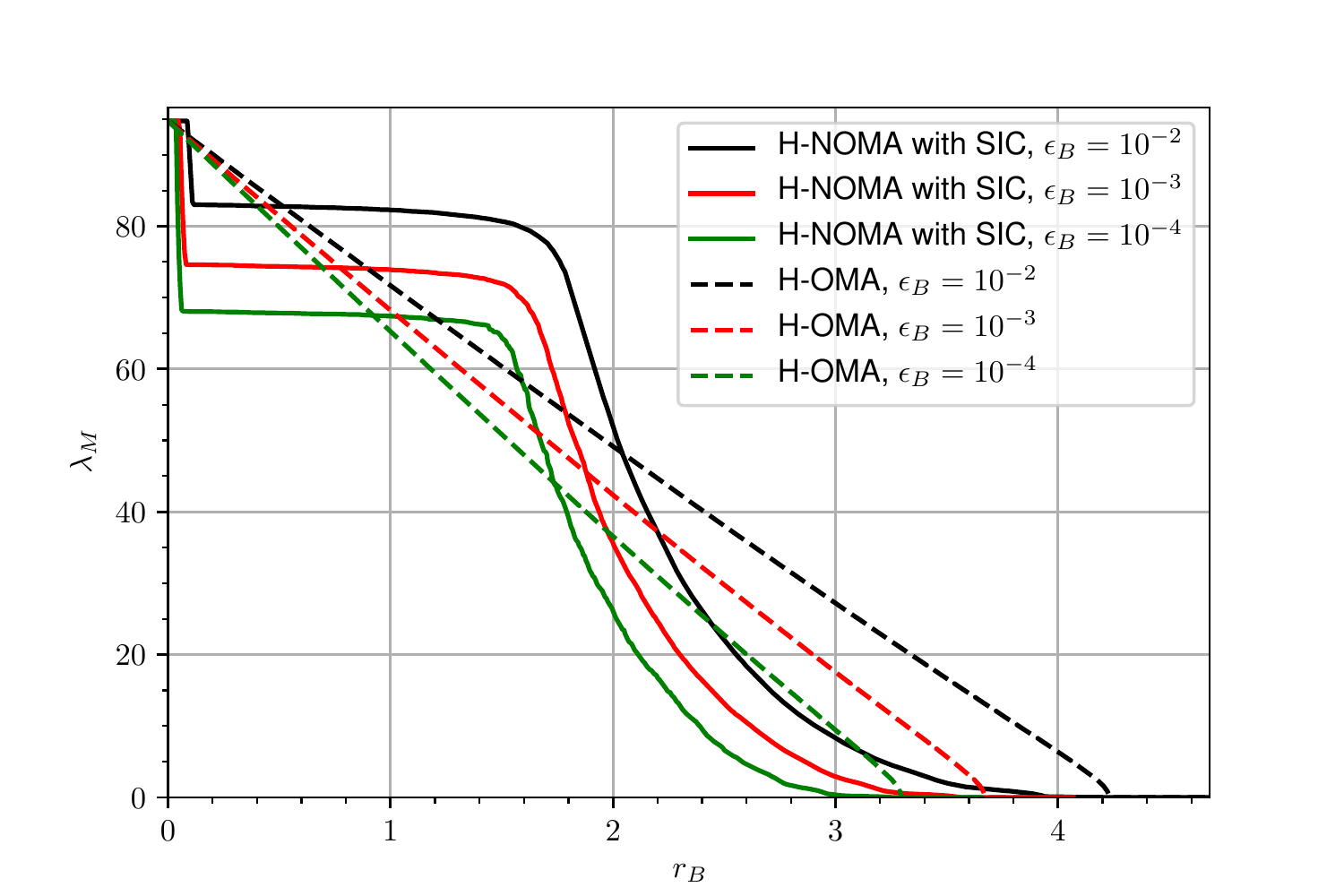}
  \caption{Arrival rates $\lambda_M^{\text{orth}}$ and $\lambda_M^{\text{non-orth}}$ for mMTC traffic under H-OMA and H-NOMA, respectively, as a function of the eMBB rate $r_B$ for $\epsilon_B\in\{10^{-2},10^{-3},10^{-4}\}$. The parameters are $\Gamma_M=5$ dB, $\Gamma_B=20$ dB, $\epsilon_M=10^{-1}$, and $r_M = 0.04$.}
 \label{fig:embb_mmtc_fig3_GamMdB_5_GamBdB_20_rM_0_0400_epsM_1.pdf}
\end{figure}

In Figs.~\ref{fig:embb_mmtc_fig2_GamMdB_5_rM_0_0400_epsM_1_epsB_3} and  \ref{fig:embb_mmtc_fig3_GamMdB_5_GamBdB_20_rM_0_0400_epsM_1.pdf}, we plot the supported mMTC arrival rate $\lambda_M$ as a function of $r_B$ for different values of the eMBB SNR $\Gamma_B$ and $\epsilon_B=10^{-3}$, and for different eMBB reliability levels $\epsilon_B$ and $\Gamma_B=20$ dB, respectively. This figures allows us to assess the impact of the average eMBB gain $\Gamma_B$ and of the eMBB reliability $\epsilon_B$ on the operation of the system in the three regimes identified above and on relative performance of orthogonal and non-orthogonal slicing, as discussed next. 

As it pertains to three regimes, we observe from Fig.~\ref{fig:embb_mmtc_fig2_GamMdB_5_rM_0_0400_epsM_1_epsB_3} that the rate at which the transition from the second to the third regime occurs does not change as the eMBB average channel gain $\Gamma_B$ is increased from $20\text{ dB}$ to $30\text{ dB}$. 
On the contrary, the $r_B$ value corresponding to the transition from the first to  the second regime becomes larger with this increase in $\Gamma_B$. This increase, in fact, allows the eMBB transmission to be decoded earlier in the SIC process for a larger set of values of $r_B$. From Fig.~\ref{fig:embb_mmtc_fig3_GamMdB_5_GamBdB_20_rM_0_0400_epsM_1.pdf}, we observe that the eMBB error probability constraint significantly affects the supported mMTC arrival rate in the second and third regimes. In fact, in these case, the higher eMBB transmission power required to ensure a higher reliability impairs the decoding of mMTC users via interference.

%Intuitively, as $\Gamma_B$ is  increased, the upper bound on $G_B^\text{tar}$ on the right-hand side of \eqref{eq:BetaFeta3} also increases. However, it is not always beneficial to increase $G_B^\text{tar}$ up to the point in which \eqref{eq:BetaFeta3} is satisfied with equality. In particular, increasing $G_B^\text{tar}$ above a certain value in the third regime implies that more mMTC devices cannot be decoded because of the increased eMBB interference. This in turn will also result in more mMTC interference and, in some cases, reduce the number of supported mMTC devices.  

We now elaborate on the comparison between orthogonal and non-orthogonal slicing. The presented figures emphasize the fact that there are points in the rate region ($r_B$,$\lambda_M$) that can be attained by non-orthogonal slicing and not by orthogonal slicing, and vice versa. Specifically, non-orthogonal slicing is seen to be beneficial when $r_B$ is across the second and the third regimes, especially for not too large reliability levels $\epsilon_B$ (see Fig. \ref{fig:embb_mmtc_fig2_GamMdB_5_rM_0_0400_epsM_1_epsB_3}). For such values, the eMBB rate is large, and yet low enough not to hamper the decoding of the mMTC users. Once again, reliability diversity is crucial to ensure the effectiveness of non-orthogonal slicing. 
In contrast, for large values of the rate $r_B$, when non-orthogonal slicing is deeply in the third operating regime, orthogonal slicing is always superior. This is because in this regime the performance is limited by the interference caused by eMBB users.

%For non-orthogonal slicing, it  seems favorable to operate at the transition between the first and second regime, which we have denoted by $r_B^{(1)}$ in Fig.~\ref{fig:embb_mmtc_fig1_GamMdB_5_GamBdB_25_rM_0_0400_epsM_1_epsB_3}, or at the transition between the second and third $r_B$-regime, which we have denoted by $r_B^{(2)}$ in Fig.~\ref{fig:embb_mmtc_fig1_GamMdB_5_GamBdB_25_rM_0_0400_epsM_1_epsB_3}. 


\section{Discussion, Conclusions and Future Work}
\label{sec:DiscussionGeneralSlicing}

\subsection{Discussion and Conclusions}\label{sec:conclusion}
In this work, we have presented a communication-theoretic model that enables the investigation of the fundamental trade-offs associated with the sharing of the wireless resources among the three 5G traffic types, namely eMBB, mMTC and URLLC. Albeit simple, the model accounts for the differences among the services in reliability, latency, and number of supported devices. 
Specifically, we have considered the slicing of resources among the services in the uplink over a shared multiple access resource. We have utilized the term ``slicing'' in order to emphasize the heterogeneous performance requirements that need to be satisfied for each service as well as the performance isolation among services. Two slicing paradigms have been investigated, orthogonal and non-orthogonal and the respective transmission schemes H-OMA and H-NOMA, where the latter is inherently possible only in shared wireless channels. 

We have applied the model to the study of the slicing for two services in two different cases: \emph{(i)} eMBB and URLLC and \emph{(ii)} eMBB and mMTC. 
In both cases, we have shown that, in order to be effective, the design of non-orthogonal slicing solutions must be guided by reliability diversity.
For the case of eMBB-URLLC coexistence, reliability diversity dictates that, in H-NOMA with SIC, the URLLC device should be decoded first, as its decoding cannot depend on the decoding of eMBB, whose reliability and latency requirements are much looser compared to URLLC. The implications of reliability diversity are more subtle in the case of non-orthogonal slicing between eMBB and mMTC. In this case, considering the fact that the number of active mMTC devices is large with high probability, it is natural to introduce a reliability metric that accounts for the fraction of correctly decoded transmissions. The analysis demonstrated that there are regimes in which the decoding of eMBB should be performed after the decoding of one or multiple mMTC devices in order to benefit from non-orthogonal slicing.
% However, there is no specific mMTC device that should always be decoded before the eMBB device, as in that case that mMTC device would achieve reliability higher than eMBB. Hence, the set of mMTC devices that need to be decoded before the eMBB device is random and, on average, each of them attains the reliability required by the mMTC devices.

Our numerical results show that there are regimes in which H-NOMA is advantageous over H-OMA and vice versa. In the case of eMBB-URLLC, H-NOMA with SIC is always beneficial when the eMBB rate is very large. 
In the case of eMBB-mMTC, non-orthogonal slicing is beneficial when the eMBB rate takes values that are small enough not to hamper the decoding of the mMTC devices.
\subsection{Generalizations and Future Work} \label{sec:extensions}

The analysis presented in this paper is based on some simplifying assumptions. 
However, the basic model and the methodology developed here can be extended to more general models and other operation regimes, as briefly discussed here.

Starting with eMBB-URLLC coexistence, one could devise another H-NOMA scheme, where eMBB and URLLC users are allowed to access partially non-orthogonal resources, so that only a subset of frequency channels potentially occupied by URLLC traffic may be interfered by eMBB transmissions. Another direct generalization is to assume that the minislots are pre-allocated to different URLLC devices. Recall that, when all transmissions are made by the same URLLC device, the block fading model dictates that either all or none of the transmissions in a minislot are decoded correctly. If each URLLC transmission is carried out by a different device, then then error decoding events are independent across the minislots. As a more involved extensions of the model, one may consider the impact of frequency diversity also for eMBB traffic, and the performance under alternative decoding strategies, such as treating interference as noise.

%the case in which eMBB dcoding treats the URLLC signals as noise and attempts decoding even if the URLLC signals have not been decoded or the minislots have not been erased (as in puncturing). A limitation of our model is that eMBB transmissions does not use coding and power distribution across different frequency channels. If an eMBB transmission spans multiple frequency channels, then  decoding is successful if the mutual information summed over all slots exceeds a threshold related to the rate. In this case, the outage probability depends on the realizations of all the $\{G_{B,f}\}$.

As for the coexistence of mMTC and eMBB services, an interesting extension is to allow multiple channels for mMTC traffic. In particular, mMTC devices may be allowed to use frequency hopping, and the number of allocated frequency channels may depend on the reliability requirements. Another aspect that deserves study is the impact of the arrival process, which here has been assumed to be Poisson. Namely, the higher burstiness of the arrival process can potentially improve the gain that one can obtain with non-orthogonal slicing. 
Finally, following the approach in NB-IoT systems, the transmission of a single mMTC device may consist  of replicas of the same packet in multiple time slots. This makes the non-orthogonal slicing of mMTC and eMBB even more relevant, as it is not feasible to reserve resources exclusively for replicas of  packets generated by sporadically active mMTC devices.

\appendices

\section{Lower Bound on the URLLC rate $r_U$}
\label{sec:LBr_U}

By setting $\Pr(E_U)=\epsilon_U$, we can upper bound (\ref{r_Udecoding}) as
\begin{IEEEeqnarray}{rCl}
\epsilon_U &=&\Pr\lefto[\sum_{f=1}^{F_U} \log_2 \left(1+\frac{G_{U,f}}{1+\delta_f G^{\mathrm{tar}}_{B}}\right) < F_U r_U \right ] \label{eq:r_U_Markov_1} \\
&{=}&\Pr \lefto[ \prod_{f=1}^{F_U} \left(1+\frac{G_{U,f}}{1+\delta_f G^{\mathrm{tar}}_{B}}\right)^{-t} \geq 2^{-r_UF_Ut}\right]  \label{eq:r_U_Markov_2}\\
&{\leq}& \frac{\mathbb{E}\lefto[ \prod_{f=1}^{F_U} \left(1+\frac{G_{U,f}}{1+\delta_f G^{\mathrm{tar}}_{B}}\right)^{-t} \right]}{2^{-r_UF_Ut}} \label{eq:r_U_Markov_3}\\
&{=}&\frac{\mathbb{E} \lefto[ \left(1+\frac{G_{U,1}}{1+\delta_1 G^{\mathrm{tar}}_{B}}\right)^{-t} \right]^{F_U}}{2^{-r_UF_Ut}}. \label{eq:r_U_Markov}
% % \sum_{f=1}^{F_U} \log_2 \left(1+\frac{G_{U,f}}{1+\delta_f G^{\mathrm{tar}}_{B}} \right) < F_U r_U \right  )
\end{IEEEeqnarray}
{We obtained~\eqref{eq:r_U_Markov_1} by multiplying both terms in the inequality by $-t$ and then by exponentiating them;~\eqref{eq:r_U_Markov_2} follows from Markov inequality; and~\eqref{eq:r_U_Markov_3} holds because $\{G_{U,f}\}$ and $\{\delta_f\}$, $f\in \{ 1,\dots, F_U\}$ are \iid, and hence we can set $f=1$ in~\eqref{eq:r_U_Markov} without loss of generality.} 
The inequality in (\ref{eq:r_U_Markov}) can be rewritten as
\begin{align}
r_U &\geq \frac{1}{tF_U}\log_2 \epsilon_U -\frac{1}{t} \log_2 \mathbb{E}\mathopen{} \left[ \left(1+\frac{G_{U,1}}{1+\delta_1 G^{\mathrm{tar}}_{B}}\right)^{-t} \right] \label{eq:r_U_CorrectEt} 
{>} \frac{1}{tF_U}\log_2 \epsilon_U -\frac{1}{t} \log_2 \mathbb{E}\mathopen{} \left[ \left(1+\frac{G_{U,1}}{1+G^{\mathrm{tar}}_{B}}\right)^{-t} \right] 
% \nonumber \\
% &\stackrel{(e)}{\geq} \frac{1}{tF_U}\log_2 \epsilon_U + \log_2 \left(1+\frac{\Gamma_U}{1+G^{\mathrm{tar}}_{B}} \right) - \frac{1}{t} \log_2 \Delta_t \label{eq:r_U_Delta_t}
\end{align}
where the strict lower bound follows by assuming that the eMBB interference is always present, i.e., $\Pr(\delta_1)=1$. The expectation in (\ref{eq:r_U_CorrectEt}) can be calculated by a Monte Carlo simulation and the value of $t\geq 0$ in~\eqref{eq:r_U_CorrectEt} is chosen such as to maximize the lower bound. 

\section{Bounds for Non-Orthogonal Slicing for eMBB and mMTC}
\label{sec:AppeMBBmMTC}

\subsubsection{{A first upper bound}} {The idea behind the bound is as follows: if the eMBB decoding fails, then the decoding of all mMTC devices for which $G_{[m]}/(1+G^{\text{tar}}_B) \leq 2^{r_M}-1$ must also fail}. 
We obtain next a lower bound on the eMBB error probability, which will give us the desired upper bound on $\lambda_M$,  by assuming that all mMTC devices that do not satisfy $G_{[m]}/(1+G^{\text{tar}}_B) \leq 2^{r_M}-1$ are decoded correctly and cancelled before eMBB decoding, and that the remaining ones, which are not decoded correctly, cause interference to the eMBB.
This yields
\begin{IEEEeqnarray}{rCl}
\Pr(E_B)
    &\geq& 1-a_B + a_B  \prBigg{\frac{G^{\text{tar}}_{B}}{1+\sum_{m=1}^{A_M} G_{[m]} \indi{\frac{G_{[m]}}{1+G^{\text{tar}}_B} \leq 2^{r_M} - 1}} \leq 2^{r_{B}}-1}\IEEEeqnarraynumspace\\
     &\geq& 1-a_B + a_B \prBigg{\frac{G^{\text{tar}}_B}{2^{r_{B}}-1} - 1 \leq \underbrace{\sum_{m=1}^{A_M} G_{[m]} \indi{\frac{G_{[m]}}{1+G^{\text{tar}}_B} \leq 2^{r_M} - 1}}_{=\chi}}.\IEEEeqnarraynumspace \label{eq:UpperLambda} 
\end{IEEEeqnarray}
{Here, $\indi{\mathcal{A}}$ is the indicator function of the event $\mathcal{A}$.}
The random variable $\chi$ in~\eqref{eq:UpperLambda} is the sum of a Poisson-distributed number of truncated, exponential-distributed random variables, which allows for an efficient numerical evaluation of the probability term in~\eqref{eq:UpperLambda}.
Next, we set the right-hand side of~\eqref{eq:UpperLambda} equal to $\epsilon_B$, and find the values of $G_B^{\text{min}}$ and $G_B^{\text{tar}}$ that result in the largest $\lambda_M$. 
This value is precisely the desired upper bound on $\lambda^{\text{non-orth}}_M(r_{B})$.

\subsubsection{{A lower bound and an alternative upper bound}} 
We upper-bound  the eMBB error probability by considering the following suboptimal decoding scheme. We force the decoder to always decode the eMBB first and subsequently decode the mMTC devices. The maximal supported arrival rate with this modified decoder is clearly a lower bound on $\lambda_M^{\text{non-orth}}(r_{B})$. 
{While deriving this bound, we shall derive as by-product also an alternative upper bound on $\lambda_M^{\text{non-orth}}(r_{B})$.}

As already mentioned, decoding the eMBB as the first device results in an upper bound on the eMBB error probability. Mathematically,
\begin{IEEEeqnarray}{rCl}
	\Pr(E_B)
    &\leq& 1 - a_B + a_B\pr{ \frac{G^{\text{tar}}_B}{1 + \sum_{m=1}^{A_M} G_{[m]}}\leq 2^{r_{B}}-1 }\IEEEeqnarraynumspace\\
    &=& 1 - a_B + a_B\pr{ \frac{G^{\text{tar}}_B}{2^{r_{B}} - 1} - 1 \leq \sum_{m=1}^{A_M} G_{[m]}}.\label{eq:erlang}
\end{IEEEeqnarray}
Here, the random variable $\sum_{m=1}^{A_M} G_{[m]}$ follows an Erlang distribution. 
It will turn out convenient to denote by $q_B(r_B, \lambda_M)$ the right-hand side of~\eqref{eq:erlang}.
Let now $\mathcal{E}_M^{\text{orth}}(\lambda_M)$ be the mMTC error probability as a function of mMTC arrival rate $\lambda_M$ in the absence of the eMBB device. 
By the law of total probability, the mMTC error probability when the eMBB is present can be upper- and lower-bounded as
\begin{IEEEeqnarray}{rCl}\label{eq:error_ub_lb}
	\mathcal{E}_M^{\text{orth}}(\lambda_M) 
     \leq \pr{E_M}= 1-\frac{\E{D_M}}{\lambda_M}
    \leq \mathcal{E}_M^{\text{orth}}(\lambda_M) + \epsilon_B.
\end{IEEEeqnarray}
Set now 
\begin{IEEEeqnarray}{rCl}\label{eq:lambda_ub_lb}
	\lambda_{M,\text{lb}} &=& \max\left\{ \lambda_M \geq 0: \mathcal{E}_M^{\text{orth}}(\lambda_M)  \leq \epsilon_M-\epsilon_B \right\}\\
    \lambda_{M,\text{ub}} &=& \max\left\{ \lambda_M \geq 0: \mathcal{E}_M^{\text{orth}}(\lambda_M) \leq \epsilon_M \right\}.
\end{IEEEeqnarray}
In words, these are the largest arrival rates for which the right-hand side and the left-hand side of~\eqref{eq:error_ub_lb} are smaller than $\epsilon_M$, respectively.
Furthermore, let $r_B^{\text{low}}$ be given by
\begin{IEEEeqnarray}{rCl}
  r_B^{\text{low}} &=& \max\mathopen{}\Big\{ r_B\geq 0:  q_B(r_B, \lambda_{M,\text{lb}}) \leq \epsilon_B \Big\}.\IEEEeqnarraynumspace
\end{IEEEeqnarray}
It follows that the pair $(r_B^{\text{low}},\lambda_{M,\text{lb}})$ is achievable. Furthermore, we have that
\begin{IEEEeqnarray}{rCl} \label{eq:LowerUpperLambda}
  \lambda_{M,\text{lb}} \leq \lambda_{M}^{\text{non-orth}}(r_{B}) \leq \lambda_{M,\text{ub}}
\end{IEEEeqnarray}
for all $r_B \in [0,r_B^{\text{low}}]$. 
When $r_B\in (r_B^{\text{low}},r_{B}^{\text{orth}}]$, we obtain a lower bound on $\lambda_M^{\text{non-orth}}(r_B)$ by finding the largest value of $\lambda_M$ for which  $q_B(r_B,\lambda_M)\leq \epsilon_M$ and by taking the smallest between this value and $\lambda_{M,\text{lb}}$.
We conclude by noting that the upper bound on $\lambda_{M}^{\text{non-orth}}(r_{B})$ on the right-hand-side of 
(\ref{eq:LowerUpperLambda}), which holds for all $r_B$, can be combined with the upper bound resulting from~\eqref{eq:UpperLambda} to tighten it when $r_B$ is small.

\bibliographystyle{IEEEtran}

\end{document}